\documentclass[12pt]{iopart}

\usepackage{epsfig}
\usepackage{amsmath}
\usepackage{graphicx}
\usepackage{bm}
\usepackage[T1]{fontenc}
\usepackage[latin1]{inputenc}
\usepackage{times}
\usepackage{amssymb}

\begin{document}

\title{Controlling Light Through Optical Disordered Media~: Transmission Matrix Approach}

\author{S. M. Popoff, G. Lerosey, M. Fink, A. C. Boccara, S. Gigan}

\address{Institut Langevin, ESPCI ParisTech, CNRS UMR 7587, 10 rue Vauquelin, 75231 Paris, France.}

\ead{sylvain.gigan@espci.fr}

\date{\today}

\begin{abstract}
We experimentally measure the monochromatic transmission matrix (TM) of an optical multiple scattering medium using a spatial light modulator together with a phase-shifting interferometry measurement method. The TM contains all information needed to shape the scattered output field at will or to detect an image through the medium. We confront theory and experiment for these applications and we study the effect of noise on the reconstruction method. We also extracted from the TM informations about the statistical properties of the medium and the light transport whitin it. In particular, we are able to isolate the contributions of the Memory Effect (ME) and measure its attenuation length.

\end{abstract}

\submitto{\NJP}	
\maketitle

\section*{Introduction}

For a long time, wave propagation was only studied in the simple cases of homogeneous media or simple scattering were the trajectories of waves can be exactly determined~\cite{goodman2005introduction}. Nevertheless, the effect of the multiple elastic scattering on propagation of waves is complex but deterministic, and information conveyed inside such media is shuffled but conserved~\cite{PhysRevE.67.036621}. Within the last decades, the study of complex and random media has been an interdisciplinary subject of interest, ranging from solid state physics to optics via acoustics, electromagnetism, telecommunication and seismology~\cite{ishimaru1978wave,sheng1995introduction,tourin2000multiple,lee1985disordered,margerin1998radiative}.\\
Recent works in optics have shown that it is possible to achieve wavefront shaping beyond a complex media despite multiple scattering~\cite{vellekoop2007focusing,yaqoob2008optical} and even to take advantage of the randomness of such media to increase resolution~\cite{vellekoop2010exploiting} by reducing the size of the focal spots. These works amount to performing phase conjugation. Nevertheless, in acoustics~\cite{prada1994time}, seismic~\cite{goupillaud1961approach}, electromagnetics or telecommunication~\cite{introPaulraj}, other operations are done to transmit information through  a random propagation systems. These techniques assume that the propagation between input and output is characterized in amplitude and phase, which is not trivial in optics.\\
We recently introduced a method to experimentally measure the monochromatic transmission matrix (TM) of a complex media~\cite{popoff2010measuring} and presented an application to image transmission through such media~\cite{popoff2010image}. In the present paper, we will develop the TM model for optical systems, in particular for multiple scattering media and
detail the experimental acquisition method. We will show that the TM can be exploited for different applications. The most straightforward ones are focusing and image detection using well-established operators used for inverse problems in various domains (telecommunication~\cite{introPaulraj}, seismology~\cite{goupillaud1961approach}, optical tomography~\cite{arridge1999optical,maire2009experimental}, acoustics~\cite{FinkPhysTo97,montaldo2004real} and electromagnetics~\cite{PhysRevLett.92.193904}). We will confront theory and experiment for applications and study the statistical properties of the TM. We will in particular study the effect of the ballistic contributions on the statistics of the TM and its consequences on image transmission. We will finally study how the short range correlation of the Memory Effect~\cite{PhysRevLett.61.2328} (ME) affects the TM and how to quantify its correlation length with the knowledge of the TM.

\section{Matrix Model and Acquisition}
\label{SecLinear}

\subsection{Modeling Transmission Through a Linear Optical System}
\label{model}

In free space, a plane wave propagates without beeing modified and its \textbf{k}-vector is conserved since plane waves are the eigenmodes of the free space propagation. In contrast, a plane wave illuminating a multiple scattering sample gives rise to a seemingly random output field. This output field is different for a different input \textbf{k}-vector. In order to use a scattering medium as an optical tool, one has to characterize the relation between input and output free modes. At a given wavelength, we will model such a system with a matrix linking the \textbf{k}-vector components of the output field to those of the input field. We will justify the validity and the interest of such a model.

For any linear propagation medium, the propagation of an optical wave is entirely characterized by its Green function~\cite{papas1965theory}. If the sources are linearly polarized and the observation is made on the same polarization, one can use a scalar model. The scalar Green function $G(\bm{r},\bm{r'},t,t')$ describes the influence of the optical field at position $\bm{r'}$ at time $t'$ on the optical field at position $\bm{r}$ at time $t$. For a propagation in a medium stationary over time, the time dependence of the Green function is governed by $t-t'$. For a surface $S_{\mathrm{src}}$ containing all the sources, the optical field on $\bm{r}$ at $t$ reads :

\begin{equation}
 E(\bm{r},t) = \iint\limits_{S_{\mathrm{src}}} \int_{-\infty}^{\infty} G(\bm{r},\bm{r'},t-t') E(\bm{r'},t') \,\rmd t' \,\rmd^2\bm{r'}
\end{equation}

This expression can be written in the spectral domain :

\begin{equation}
 E(\bm{r},\omega) = \iint\limits_{S_{\mathrm{src}}}  G(\bm{r},\bm{r'},\omega) E(\bm{r}',\omega)  \,\rmd^2\bm{r'}
\end{equation}

Where $E(\bm{r},\omega)$ (resp. $G(\bm{r},\bm{r'},\omega)$) is the temporal Fourier transform of $E(\bm{r',t})$ (resp. $G(\bm{r},\bm{r'},t)$ ) at the pulsation $\omega$. We want to characterize the relation between the optical field on the surface $S_{\mathrm{src}}$ containing the sources and the optical field on the elements of an observation surface $S_{\mathrm{obs}}$. Experimentally, sources and receptors have a finite size. We note $E^{\mathrm{out}}_m = \int_{S^{\mathrm{out}}_m}E(\bm{r})\,\rmd\bm{r}$ the average optical field on the $m^{th}$ receptor of surface $S^{\mathrm{out}}_m \subset S_{\mathrm{obs}}$. In a similar way, we note $E^{\mathrm{in}}_n = \int_{S^{\mathrm{in}}_n}E(\bm{r})\,\rmd\bm{r}$ the average optical field on the $n^{th}$ controlled area of surface $S^{\mathrm{in}}_n \subset S_{\mathrm{src}}$. We write $N$ and $M$ respectively the number of sources and receptors.

Therefore, we can define the mesoscopic TM of an optical system for a given wavelength as the matrix $K$ of the complex coefficients $k_{mn}$ connecting the optical field (in amplitude and phase) at the $m^{th}$ output element to the one at the $n^{th}$ input element element. Thus, we have :

\begin{equation}
 E^{\mathrm{out}}_m = \sum_n^N{k_{mn}E^{\mathrm{in}}_n}
\label{eqE}
\end{equation}

In essence, the TM gives the relationship between input and output pixels at a given frequency, notwithstanding the complexity of the propagation for a stationary medium. Along this paper, we will perform the Singular Value Decomposition (SVD) of TMs to study their transmission properties. The SVD consists in writting~:

\begin{equation}
 K = U\Sigma V^{\dag}
\label{SVD}
\end{equation}

Where $^{\dag}$ denotes the transpose conjugate. The SVD decomposes the system in independent transmission channels characterized by their transmission coefficients and by their corresponding input and output modes. $V$ is a unitary change of basis matrix linking input freemodes with transmission channel input modes of the system. $\Sigma$ is a diagonal matrix containing real and positive values called singular values of $K$ and noted $\lambda_{m}$ . These values are the square root of the energy transmission values of the transmission channels. $U$ is the unitary change of basis matrix between transmission channels output modes and output freemodes. For practical purposes, we will study the normalized singular values defined by~:

\begin{equation}
 \widetilde{\lambda}_{m} = \frac{\lambda_{m}}{\sqrt{\sum_k{\lambda^2_{k}}}}
\label{lambdanorm}
\end{equation}

The SVD is a powerful tool that gives access to the statistical distribution of the incident energy injected thought the multiple transmission channels of the system. Within this matrix model, the number of channels of the observed system is $min(N,M)$. Nevertheless, this number does not quantify the information that could simultaneously be conveyed in the medium.

Quantifying information is of high interest and this work has been done in acoustics~\cite{derode2001scattering,lemoult2009} for Time Reversal focusing through a multiple scattering medium. In those theories, the number of spatial and temporal Degrees of Freedom ($N_{\mathrm{DOF}}$) also called Number of Information Grains is introduced. These degrees of freedom represent as many independent information grains can be conveyed through the system. For a monochromatic experiment there are only spatial degrees of freedom and $N_{\mathrm{DOF}}$ is given by the number of uncorrelated element used to recover information~\cite{PhysRevE.67.036621}. This number will be different on the type of experiment considered, focusing or image detection. A focusing experiment consists in finding the optimal input wavefront to send to maximize energy at a desired position (or a combination of positions) on the output of the sample. An image detection experiment consists in retrieving an image sent at the input of the system by measuring the output complex field. For focusing experiments $N_{\mathrm{DOF}}$ is given by the number of independent input segments, which corresponds to the number of degrees of freedom the experimentalist can control. Conversely, for detection experiments, the experimentalist cannot access more than the number of independent output elements recorded. In the general case, we have $N_{\mathrm{DOF}}  \leq N$ for focusing and $N_{\mathrm{DOF}} \leq M$. We choose to fix the input (resp. output) pixel size at the input (resp. output) coherence length of the system. For this ideal sampling, $N_{\mathrm{DOF}}$ is given by the input or output dimension of the TM according to the type of experiment : $N_{\mathrm{DOF}} = N$ for focusing and $N_{\mathrm{DOF}} = M$ for detection.

\subsection{Experimental Setup}
\label{expsetup}

To measure the TM of an optical system, one needs a way to control the incident light and to record the complex optical field. The control is now possible thanks to the emergence of spatial light modulators (SLM) and the output complex field can be measured using an interferometric method detailed in section~\ref{AcqMethod}.
We want to study an optical system governed by multiple scattering in which the propagation function can not be \textit{a priori} calculated. The material chosen is Zinc-Oxyde, a white pigment commonly used for industrial paints. The sample is an opaque 80 $\pm$ 25 $\mu m$ thick deposit of a ZnO powder (Sigma-Aldrich 96479) with a measured mean free path of $6 \pm 2$ $\mu$m on a standard microscope glass slide. The thickness of the sample being one order of magnitude higher than the mean free path, light transport inside the sample is in the multiple scattering regime. To generate the incident wavefront, the beam from a diode-pumped solid-state single longitudinal mode laser source at 532 nm (Laser Quantum Torus) is expanded and spatially modulated by a liquid crystal SLM (Holoeye LC-R 2500). Using a polarizer and an analyzer, we choose an appropriate combination of polarizations before and after the SLM to achieve up to a $2\pi$ phase modulation depth with less than 10$\%$ residual amplitude modulation. Only a part of the SLM is modulated and the rest of the surface is used to generate a static reference (details in \ref{AcqMethod}). The surface of the SLM is then imaged on the pupil of a 20x objective with a numerical aperture (NA) of 0.5. Thus a pixel of the SLM matches a wave vector at the entrance of the scattering medium. The beam is focused at one side of the sample. We image the output pattern 0.3 mm from the opposite surface of the sample through a 40x objective (NA = 0.85) onto a 10-bit CCD camera (AVT Dolphin F-145B). We do not directly image the surface of the material to mitigate the effects of the possible presence of ballistic light. This point is detailed in section~\ref{BallSec}.

The duration of a typical measurement of the TM is of a few minutes, much less than the typical decorrelation time of the medium (more than an hour). From now on, we will ignore any time dependence of the medium. It is an important prerequisite for the measurement and exploitation of the TM.

\begin{figure}[ht]
\center
\includegraphics[width=0.65\textwidth]{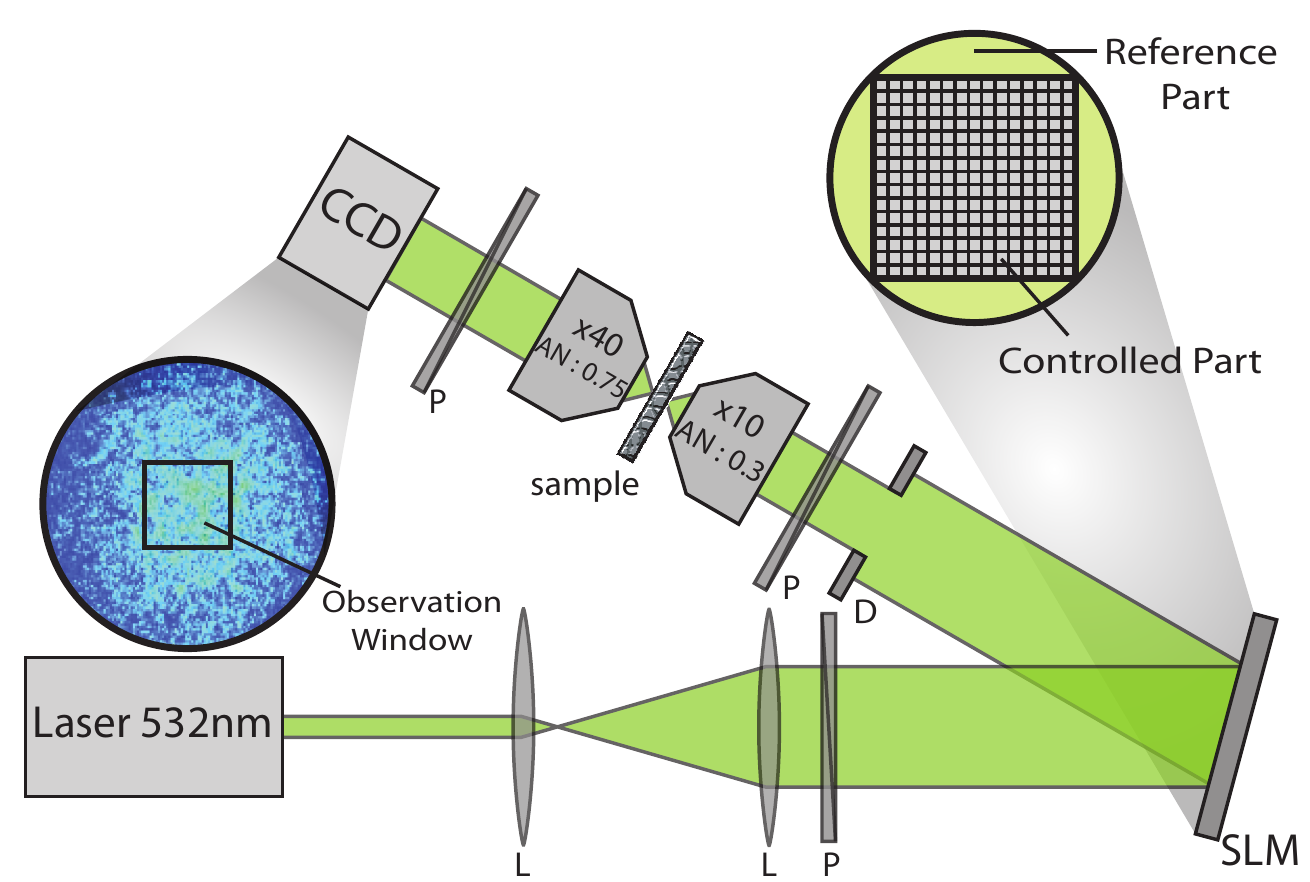}
\caption{(color online) Schematic of the apparatus. The 532 nm laser is expanded and reflected off a SLM. Polarization optics select a phase only modulation mode. The modulated beam is focused on the multiple-scattering sample and the output intensity speckle pattern is imaged by a CCD-camera. L, lens. P, polarizer. D, diaphragm.}
\label{Setup}
\end{figure}

\subsection{Acquisition Method}
\label{AcqMethod}

Measuring a TM in optics raises several difficulties coming from the impossibility of having direct access to the phase of the optical field. Generally, one can have access to the complex optical field using interferences, for instance with a known wavefront and a full field phase-shifting interferometry method~\cite{zhang1998three}. For any input vector, if the reference phase is shifted by a constant $\alpha$, the intensity in the $m^{th}$ output mode is given by~:

\begin{eqnarray}
  I_m^{\alpha} &=& |E^{\mathrm{out}}_m|^2 = |s_m + \sum_n^N{\rme^{\rmi \alpha}k_{mn}E^{\mathrm{in}}_n}|^2 \\ \nonumber
  &= &|s_m|^2+|\sum_n^N{\rme^{\rmi\alpha}k_{mn}E^{\mathrm{in}}_n|^2}
 +2\Re{\left(\rme^{\rmi\alpha}\overline{s}_m \sum_n^N{k_{mn}E^{\mathrm{in}}_n }\right)} \nonumber
\label{equI}
\end{eqnarray}

where $s_m$ is the complex amplitude of the optical field used as reference in the $m^{th}$ output mode.

Thus, if we inject the $n^{th}$ input mode and measure $I^{0}_m$,$I^{\frac{\pi}{2}}_m$,$I^{\pi}_m$ and $I^{\frac{3\pi}{2}}_m$, respectively the intensities in the $m^{th}$ outgoing mode for $\alpha=0$, $\pi/2$, $\pi$ and $3\pi/2$ gives :	
\begin{eqnarray}
 \frac{\left(I^{0}_m-I^{\pi}_m\right)}{4}+ i \frac{\left(I^{\frac{3\pi}{2}}_m-I^{\frac{\pi}{2}}_m\right)}{4} = \bar{s}_m k_{mn}
\label{4phi}
\end{eqnarray}

Commercial SLMs cannot control independently phase and amplitude of an incident optical field. Our setup uses a phase-only modulator since the phase is the most important parameter to control in wavefront shaping~\cite{love1997wave}. A recent method~\cite{van2008spatial} allows phase and amplitude modulation with similar commercial modulators, nevertheless, we use in this work a phase-only configuration to keep the setup as simple as possible.

Controlling only the phase forbids us to switch off or modulate in amplitude an incident mode. Thus we can not directly acquire the TM in the canonical basis. We want to use a basis that utilizes every pixel with identical modulus. The Hadamard basis in which all elements are either +1 or -1 in amplitude perfectly satisfies our constraints. Another advantage is that it maximizes the intensity of the received wavefront and consequently decreases the experimental sensibility to noise~\cite{larrat2010mr}. For all Hadamard basis vectors, the intensity is measured on the basis of the pixels on the CCD camera. We obtain the TM in the input canonical basis by a unitary transformation. The observed transmission matrix $K_{\mathrm{obs}}$ is directly related to the physical one $K$  of formula~\ref{eqE} by

\begin{equation}
 K_{\mathrm{obs}} = K\times S_{\mathrm{ref}}
\label{kobs}
\end{equation}

where $S_{\mathrm{ref}}$ is a diagonal matrix of elements where $s^{\mathrm{ref}}_{mm}=s_m$ represents the static reference wavefront in amplitude and phase. 

Ideally, the reference wavefront should be a plane wave to directly have access to the $K$ matrix. In this case, all $s_{m}$ are identical and $K_{\mathrm{obs}}$ is directly proportional to $K$. However this requires the addition of a reference arm to the setup, and requires an interferometric stability. To have the simplest experimental setup and a higher stability, we modulate only 65 \% of the wavefront going into the scattering sample (this  correponds to the square inside the pupil of the microscope objective as seen in Figure \ref{Setup}), the speckle generated by the light coming from the 35 \% static part being our reference. $S_{\mathrm{ref}}$ results from the transmission of this static part and is now unknown and no longer constant along its diagonal. Nevertheless, since $S_{\mathrm{ref}}$ is stationary over time, we can measure the response of all input vectors on the $m^{th}$ output pixel as long as the reference speckle is bright enough. We will show in the next subsection that we can go back to the amplitude elements of $S_{\mathrm{ref}}$ and that the only missing information is the phase of the reference. We will quantify the effect of the reference speckle and show that neither does it impair our ability to focus or image using the TM, nor does it affect our ability to retrieve the statistical properties of the TM. 

\subsection{Multiple Scattering and Random Matrices}
\label{MSTM}

Under certain conditions, the rectangle $N$ by $M$ TM of a system dominated by multiple scattering~\cite{aubry2010singular} amounts to a random matrix of independent identically distributed entries of Gaussian statistics. We will study the properties of such matrices thanks to the Random Matrix Theory (RMT) and investigate the hypothesis for which the TM of multiple-scattering samples are well described by this formalism.

We introduced previously that the SVD of a physical matrix is a powerful tool to study the energy repartition in the different channels. For matrices of independent elements of Gaussian distribution, RMT predicts that the statistical distribution $\rho(\widetilde{\lambda})$ of the normalized singular values follows the Marcenko-Pastur Law~\cite{marcenko1967distribution}. For $\gamma = M/N$, it reads~:

\begin{eqnarray}
 \rho(\widetilde{\lambda}) = \frac{\gamma}{2\pi\widetilde{\lambda}}\sqrt{(\widetilde{\lambda}^2-\widetilde{\lambda}_{\mathrm{min}}^2)(\widetilde{\lambda}_{\mathrm{max}}^2-\widetilde{\lambda}^2)}\\
 \forall \widetilde{\lambda} \in [\widetilde{\lambda}_{\mathrm{min}}, \widetilde{\lambda}_{\mathrm{max}}] \nonumber
\label{marcenkopastur}
\end{eqnarray}

With $\widetilde{\lambda}_{\mathrm{min}} = (1-\sqrt{1/\gamma})$ the smallest singular value and $\widetilde{\lambda}_{\mathrm{max}} =  (1+\sqrt{1/\gamma})$ the maximum singular value. In particular, for $\gamma = 1$, the Marcenko-Pastur law is known as the "quarter circle law"~\cite{wigner1967random} and reads :

\begin{equation}
 \rho(\widetilde{\lambda}) = \frac{1}{\pi}\sqrt{4-\widetilde{\lambda}^2}
\label{quarterlaw}
\end{equation}

Those predictions are valid for a TM of independent elements \textit{i.e.} without any correlation between its elements.
This hypothesis can be broken by adding short range correlation but also non trivial correlations such as ones appearing when introducing the flux conservation~\cite{dorokhov1984coexistence}. Since we only experimentally record and control a small part of the field on both side of the sample, we will not be affected by this conservation condition.

Another straightforward parameter that could have an influence on the hypothesis of independent elements is the size of the segments uses as sources (on the SLM) and receptors (on the CCD). If the segment size is less than the coherence length (\textit{i.e.} the size of a speckle grain) two neighboring segments will be strongly correlated. The size of input (resp. output)  segments is set to fit the input (resp. output) coherence length. This way, the TM $K$ should respect the hypothesis of independent elements and follow the Marcenko-Pastur law.

Nevertheless the method of the acquisition introduces correlation in the TM recorded. For $M = N$, $K$ should respect the Quarter Circle law but the methods detailed in section~\ref{AcqMethod} gives only access to the observed TM matrix $K_{\mathrm{obs}} = S_\mathrm{{ref}}.K$. whose elements read~:

\begin{equation}
 k^{\mathrm{obs}}_{mn} = \sum_j^N{s_{mj}k_{jn}} = k_{mn}s_{mm}
\end{equation}

To study the singular value distribution of our TM, we have to be sure that we can go back to the precise statistics of $K$ using the observed matrix $K_{\mathrm{obs}}$.

The effect of the reference $S_{\mathrm{ref}}$ is static at the output. This means that its influence is the same for each input vectors. Calculating the standard deviation of an output elements over the input basis vectors we have :

\begin{eqnarray}
 &\sqrt{\left\langle |k^{\mathrm{obs}}_{mn}|^2\right\rangle_n} &= \sqrt{\left\langle| k_{mn}|^2\right\rangle_n}|s_{mm}| \\
 & &= \sqrt{\left\langle |k_{mn}|^2\right\rangle_{mn}}|s_{mm}| \quad \forall m \in [1,M] \nonumber
\end{eqnarray}

We made the hypothesis that each input \textbf{k}-vector gives an output speckle of the same mean intensity, it reads $\left\langle |k_{mn}|^2\right\rangle_n=\left\langle |k_{mn}|^2\right\rangle_{mn}$. This is valid in the diffusive regime as long as the illumination is homogeneous. We select for the illumination only the center part of an expanded gaussian beam. We also neglected the correlation effects that are expected to be small regarding our system, such as Memory Effect~\cite{PhysRevLett.61.2328} (see section~\ref{MemoryEffect}) and long range correlations~\cite{scheffold1998universal}. We can now define a normalized TM $K_{fil}$ whose elements $k^{\mathrm{fil}}_{mn}$ are normalized by $\sqrt{\left\langle |k^{\mathrm{obs}}_{mn}|^2\right\rangle_m}$

\begin{eqnarray}
 &k^{\mathrm{fil}}_{mn} &= \frac{k^{\mathrm{obs}}_{mn}}{\sqrt{\left\langle |k|^2\right\rangle_{mn}|s_{mm}|^2}} \\
 & &= \frac{k_{mn}}{\sqrt{\left\langle |k|^2\right\rangle_{mn}}} \frac{s_{mm}}{|s_{mm}|}
\end{eqnarray}

This matrix is not influenced by the amplitude of the reference and reads~:

\begin{equation}
K_{\mathrm{fil}} = \frac{K}{\sqrt{\left\langle |k|^2\right\rangle_{mn}}}\times S_{\phi}
\end{equation}

Where $S_{\phi}$ is a diagonal matrix of complex elements of norm 1 representing the phase contributions of the output reference speckle. In contrast with $S_{\mathrm{ref}}$, $S_{\phi}$ is a unitary matrix since $S_{\phi}S_{\phi}^{\dag} = I$ where $I$ is the identity matrix. Thus if $K = U\Sigma V^{\dag}$ then $K_{\mathrm{fil}} = U'\Sigma V^{\dag}$ where $U' = S_{\phi} \times U$ is a unitary matrix. Therefore, $K_{\mathrm{fil}}$ and $K$ have singular value decompositions with the same singular values. 

\begin{figure}[ht]
\center
\includegraphics[width=0.45\textwidth]{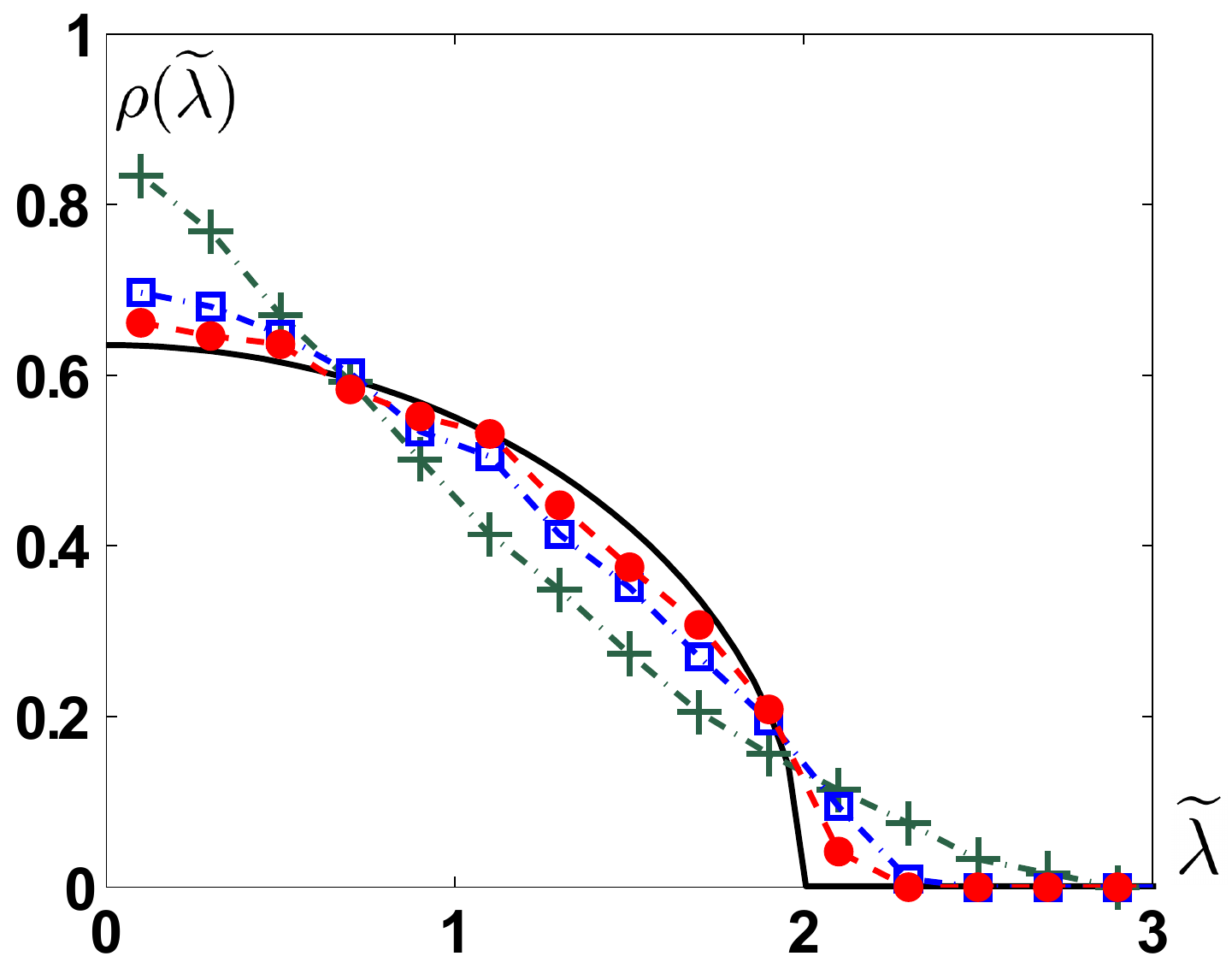}
\caption{Singular values distributions of the experimental transmission matrices obtained by averaging over 16 realizations of disorder. The solid line is the  quarter-circle law predicted by RMT.  With $+$ the observed matrix, with $\square$ the matrix filtered to remove the reference amplitude contribution and with $\circ$ the matrix obtained by filtering and removing neighboring elements to eliminate inter-element correlations. }
\label{VpSym}
\end{figure}

We show in figure~\ref{VpSym} that the statistics of the observed TM $K_{\mathrm{obs}}$ do not follow the quarter circle (formula~\ref{quarterlaw}) law whereas the singular value distribution of the filtered matrix $K_{\mathrm{fil}}$ is close to it. Remaining correlation effects are due to inter-element correlations between nearby pixels. Taking a sub-matrix of the TM with only one element out of two, its statistics finally follows correctly the expected quarter-circle law. Those results are comparable to the ones obtained for transmission matrices in acoustics~\cite{aubry2010singular}. It is to be noticed that we do not expect absorption to affect the statistical properties of the matrix measured since absorption does not induce any new correlation and since the measure is not sensitive to the total energy conservation.

This clearly demonstrates that we can directly study the properties of $K$ even if we only have access to the observed matrix $K_{\mathrm{obs}}$. We also want to point out that one has to be careful when interpreting deviation from a given prediction on SVD results. Once the matrix is measured and its singular values analyzed, one has to investigate the causes of the deviation from the Marcenko-Pastur law. Those causes could come from physical phenomenons involved in light propagation in the medium that break the hypothesis of the absence of correlation in the TM (like the apparition of ballistic contributions, see section~\ref{BallSec}) but could also be inherent to the experimental method. However, $K_{\mathrm{obs}}$ brings all informations on singular values that describe the physics of the propagation in a given setup.

\begin{figure}[ht]
\center
\includegraphics[width=0.65\textwidth]{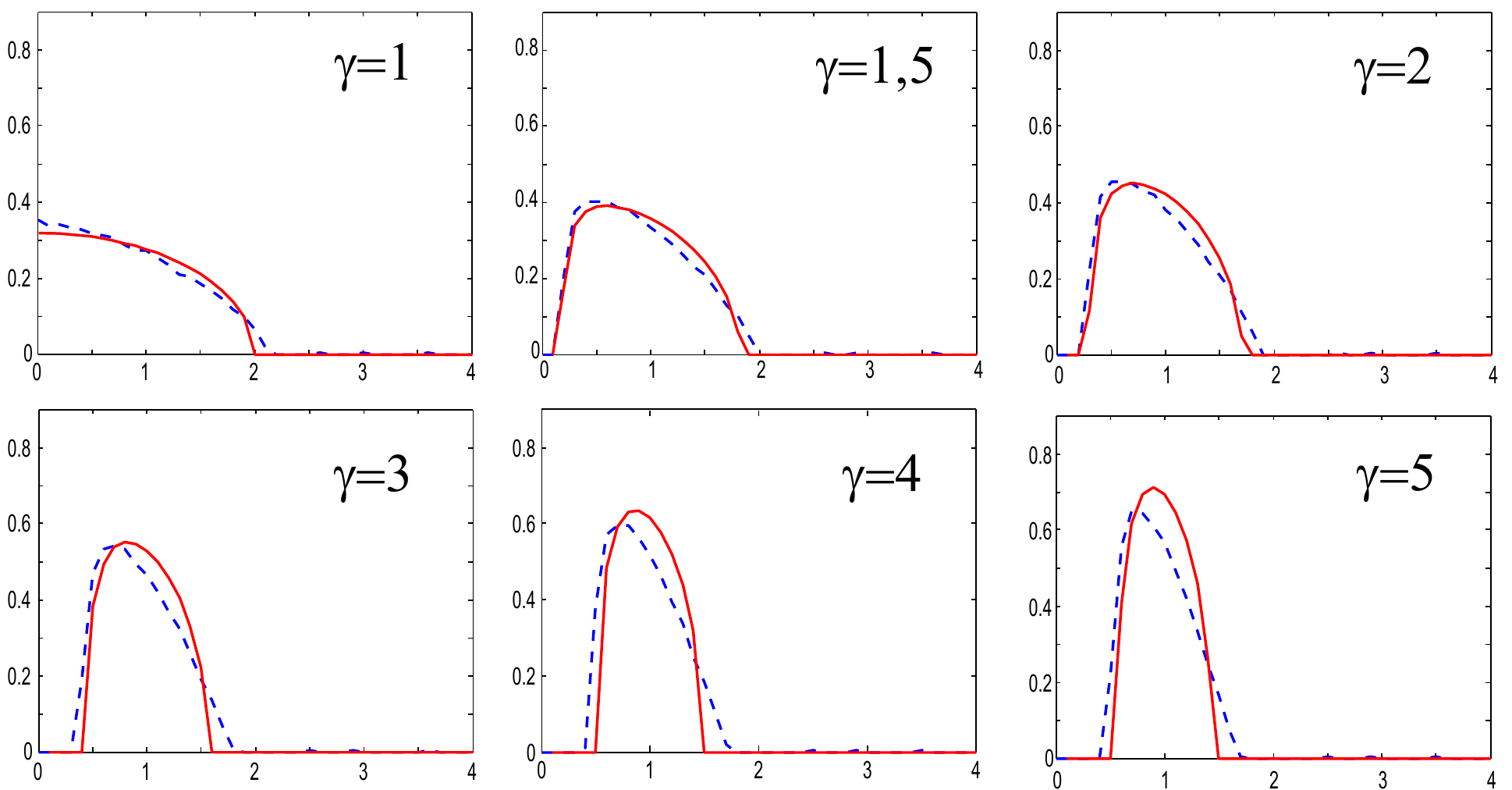}
\caption{Normalized singular value distributions for different asymmetric ratios $\gamma =$ 1, 1.5, 2, 3, 4, 5 and 6. (dash line) experimental results, (solid line) theoretical prediction from RMT. Calculations on experimental TM are made using the filtered matrix $K_{\mathrm{fil}}$.}
\label{VsAsym}
\end{figure}

So far, we compared the experimental singular value distributions for a symmetric case ($M = N$). But what would be the effect of breaking the input / output asymmetric raio $M/N$ ? In the same setup, we experimentally acquired TMs with a fixed number of controlled pixels on the SLM ($N = 1024$) and changed the number of independent output pixels recorded by the CCD ($M = \gamma N$) to achieve various ratio values ($\gamma \in \{1;1.5;2;3;4;5\}$). Normalized singular value distributions of experimental TM filtered are shown in figure~\ref{VsAsym} and follows qualitatively the RMT prediction of equation~\ref{marcenkopastur}.

We see that with increasing $\gamma$, the range of $\widetilde{\lambda}$ narrows. This effect has been measured in experiment in the acoustic field~\cite{sprik2008eigenvalue}. Physically this means that when $\gamma$ increases, all recorded channels tend to converge toward the mean global transmission value $\widetilde{\lambda} = 1$. When $M > N$, we record more independent informations than the rank of the TM and averaging effects lead to a distribution that is peaked around a unique transmission value when $M \gg N$. We notice that the experimental distributions deviate from the theoretical ones as $\gamma$ increases. To have TMs with different values of $\gamma$, we record a large TM with $\gamma$ = 11 and then create submatrices by taking lines selected randomly among the lines of the original TM. By increasing $\gamma$ we increase the probability of having neighboring pixels in the TM and then are more sensitive to correlations between nearby pixels. This modifies the statistics and decreases the number of independent output segments. This explains the deviation from the theory. Other effects that could bring correlations in certain experimental conditions are discussed is section~\ref{BallSec}.

\section{Phase Conjugation}
\label{PC}
\subsection{Phase Conjugation and Degrees of Freedom}
\label{NDOF}

It was demonstrated~\cite{PhysRevLett.92.193904,PhysRevLett.75.4206} that one can take advantage of multiple scattering to focus on tight spots, thanks to the reversibility of the wave equation.

In such an experiment, the responses to short temporal signals emitted by an array of sources at a receiver at the output of a disordered medium are recorded. Those signals, linked to the Green's functions associated to the couples  (source / receiver) are sent reversed in time. The waves generated converge naturally towards the targeted spot. The waves are focused in space and time. Phase Conjugation is the monochromatic equivalent to the previous concept named Time Reversal. Those techniques are robust methods to achieve focusing and imaging. In this chapter, we will experimentally and theoretically study the efficiency of phase conjugation on focusing.

When using Phase Conjugation for focusing, it corresponds to shape the input wavefront to put in phase the contributions of each input pixel at a desired output target. Denoting $E_{\mathrm{out}}^{\mathrm{target}}$ the output target vector, the input vector that estimates the desired output pattern $E_{\mathrm{out}}^{\mathrm{target}}$ using Phase Conjugation is given by~:

\begin{equation}
E^{\mathrm{in}} =  K^{\dag}.E_{\mathrm{out}}^{\mathrm{target}}
\label{Efoc}
\end{equation}

The effective output vector $E^{out}$ will be given by :

\begin{equation}
E^{\mathrm{out}} = K.E^{\mathrm{in}} =  K.K^{\dag}.E_{\mathrm{out}}^{\mathrm{target}}
\label{detectKKd}
\end{equation}

We will now demonstrate the dependence of the Phase Conjugation accuracy on $N_{\mathrm{DOF}}$ for the simple case of a single focusing spot. We will compare the efficiency of perfect Phase Conjugation (with amplitude modulation) and of phase only Phase Conjugation. The intensity output field resulting from the Phase Conjugation focusing on the $j^{th}$ output pixel reads~:

\begin{equation}
|s^{\mathrm{out}}_m|^2 = |\sum^{N}_l{k_{ml}k^{*}_{jl}}|^2
\end{equation}

For $m \neq j$, the average intensity of sum of the the incoherent terms reads~:

\begin{eqnarray}
\label{incoherentsum}
&\left\langle|s^{\mathrm{out}}_{m\neq j}|^2 \right\rangle &= \left\langle\sum^{N,N}_{l,l'}{k_{ml}k^*_{ml'}k^{*}_{jl}k_{jl'}} \right\rangle  \\ 
& & = \left\langle\sum^{N}_l{|k_{ml}|^2|k_{jl}|^2}\right\rangle + \left\langle\sum^{N,N}_{l\neq l'}{k_{ml}k^*_{ml'}k^{*}_{jl}k_{jl'}}\right\rangle \nonumber \\
& & = N\left\langle |k|^2\right\rangle^2 +0 \nonumber
\end{eqnarray}

For $m = j$, the average intensity of the sum of the coherent terms reads~:

\begin{eqnarray}
\label{coherentsum}
&\left\langle|s^{\mathrm{out}}_{m=j}|^2 \right\rangle &= \left\langle \left(\sum^{N}_l{|k_{jl}|^2}\right)^2\right\rangle  \\ 
& & = \left\langle \sum^{N}_{l}{|k_{jl}|^4}\right\rangle + \left\langle \sum^{N,N}_{l\neq l'}{|k_{jl}|^2|k_{jl'}|^2}\right\rangle  \nonumber \\
& &= N(N-1)\left\langle |k|^2 \right\rangle^2 + N\left\langle |k|^4 \right\rangle \nonumber \\
& &\approx N^2\left\langle |k|^2 \right\rangle^2 \quad \forall N \gg 1 \nonumber 
\end{eqnarray}

We define the energy signal to noise ratio $S\!N\!R$ for a single spot focusing as the intensity at the target point over the mean intensity elsewhere after phase conjugation. We directly have 

\begin{equation}
S\!N\!R \approx N = N_{\mathrm{DOF}} \quad \quad \forall N \gg 1
\end{equation}
\label{NDOFfoc}

Similar results are obtained considering detection instead of focusing. If $E_{\mathrm{out}}$ is the output pattern corresponding to an unknown input target mask $E_{\mathrm{in}}^{\mathrm{target}}$ to be detected, similarly to equation~\ref{detectKKd}, the reconstruction $E_{\mathrm{image}}$ of the target is given by~:

\begin{equation}
E_{\mathrm{image}} =  K^{\dag}.E_{\mathrm{out}} = K^{\dag}.K.E_{\mathrm{in}}^{\mathrm{target}}
\label{Edetect}
\end{equation}

The reconstructed pattern for detecting a single target on the $j^{th}$ input pixel reads~:

\begin{equation}
\left|s^{\mathrm{img}}_m \right|^2 = \left|\sum^{M}_l{k^{*}_{lm}k_{lj}}\right|^2
\end{equation}

And the signal to noise ratio for the reconstruction of a single pixel reads~:

\begin{equation}
S\!N\!R \approx M = N_{\mathrm{DOF}} \quad \forall M \gg 1
\label{NDOFim}
\end{equation}

\subsection{Experimental Phase Conjugation}
\label{expPC}

To perform such an ideal Phase Conjugation focusing, one has to control the amplitude and phase of the incident light. Our experimental setup uses a phase-only SLM to modulate the light. Thus the input $l^{th}$ component of the input field to display in order to focus on the $j^{th}$ output pixel reads $k^{*}_{jl}/|k^{*}_{jl}|$. The resulting output intensity pattern for a phase only experiment reads~:

\begin{equation}
\left\langle |s^{\mathrm{out}}_m|\right\rangle = \sum^{1..N}_l{k_{ml}\frac{k^{*}_{jl}}{|k^{*}_{jl}|}} = \sum^{1..N}_l|{k_{ml}|}
\end{equation}

Calculating the intensity of incoherent and coherent sums in the same way as equations~\ref{incoherentsum} and~\ref{coherentsum} we have :

\begin{eqnarray}
&\left\langle|s^{\mathrm{out}}_{m\neq j}|^2 \right\rangle &= N\left\langle |k|^2\right\rangle \\ 
&\left\langle|s^{\mathrm{out}}_{m=j}|^2 \right\rangle & \approx N^2\left\langle |k| \right\rangle^2 \quad \forall N \gg 1
\end{eqnarray}

We define $S\!N\!R_{\mathrm{exp}}$ the expected signal to noise ratio that reads~:

\begin{equation}
S\!N\!R_{\mathrm{exp}} \approx N_{\mathrm{DOF}}\frac{\left\langle|k|\right\rangle^2}{\left\langle|k|^2\right\rangle}
\end{equation}

Experimentally, for a setup and a random medium satisfying the hypothesis introduced in~\ref{MSTM} (elements of the TM independent with a Gaussian statistic), we have $\frac{\left\langle|k|\right\rangle^2}{\left\langle|k|^2\right\rangle} = \frac{\pi}{4} \approx 0.78$~\cite{vellekoop2007demixing}. Finally~:

\begin{equation}
S\!N\!R_{\mathrm{exp}} \approx \frac{\pi}{4} N_{\mathrm{DOF}}\approx 0.78 N_{\mathrm{DOF}}
\label{SNRphaseonly}
\end{equation}

To achieve such a $S\!N\!R_{\mathrm{exp}}$ using the experimental setup presented in section~\ref{expsetup} one has to remove, after measuring the TM, the reference pattern. If the reference part is still present, the intensity coming from the segments not controlled of the SLM will increase the background level. This effect (detailed in Appendix) modifies $S\!N\!R_{\mathrm{exp}}$. In our experimental conditions~:

\begin{equation}
S\!N\!R_{\mathrm{exp}} \approx  0.5 N_{\mathrm{DOF}}
\end{equation}

\begin{figure}[ht]
\center
\includegraphics[width=0.5\textwidth]{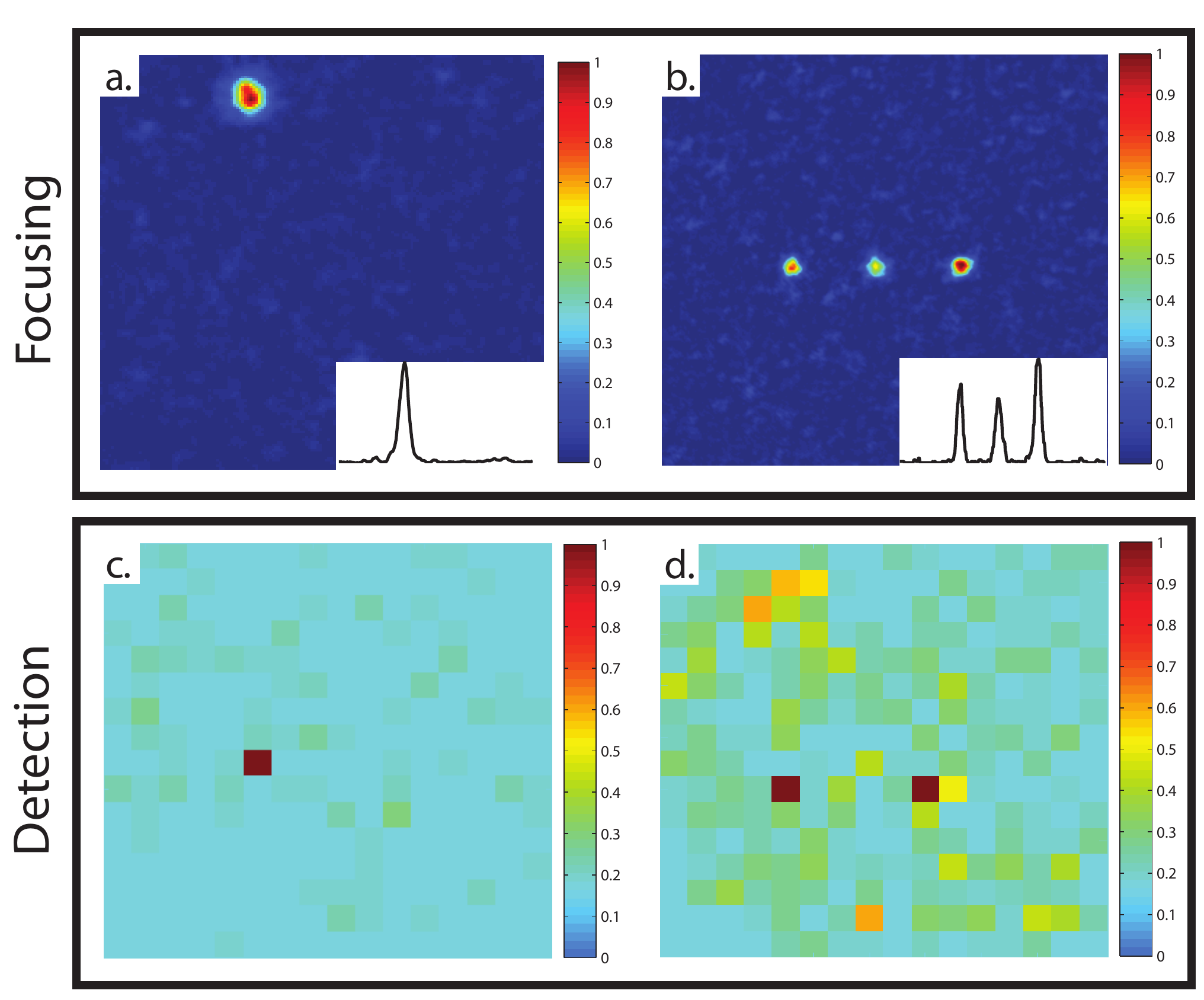}
\caption{Experimental results of focusing and detection using Phase Conjugation. Results are obtained for $N = M = 256$. We show typical output intensity pattern for a one spot target focusing \textbf{a.} and multiple target focusing \textbf{b.}. We present similar results for the detection of one point \textbf{c.} and two points \textbf{d.} before the medium. The point objects are synthesized using the SLM.}
\label{PCexp}
\end{figure}

We will now use equations~\ref{Efoc} and ~\ref{Edetect} to experimentally achieve focusing and detection. We tested Phase Conjugation for simple targets focusing and for point objects detection through the medium. The results are shown in figure~\ref{PCexp}.

\begin{figure}[ht]
\center
\includegraphics[width=0.4\textwidth]{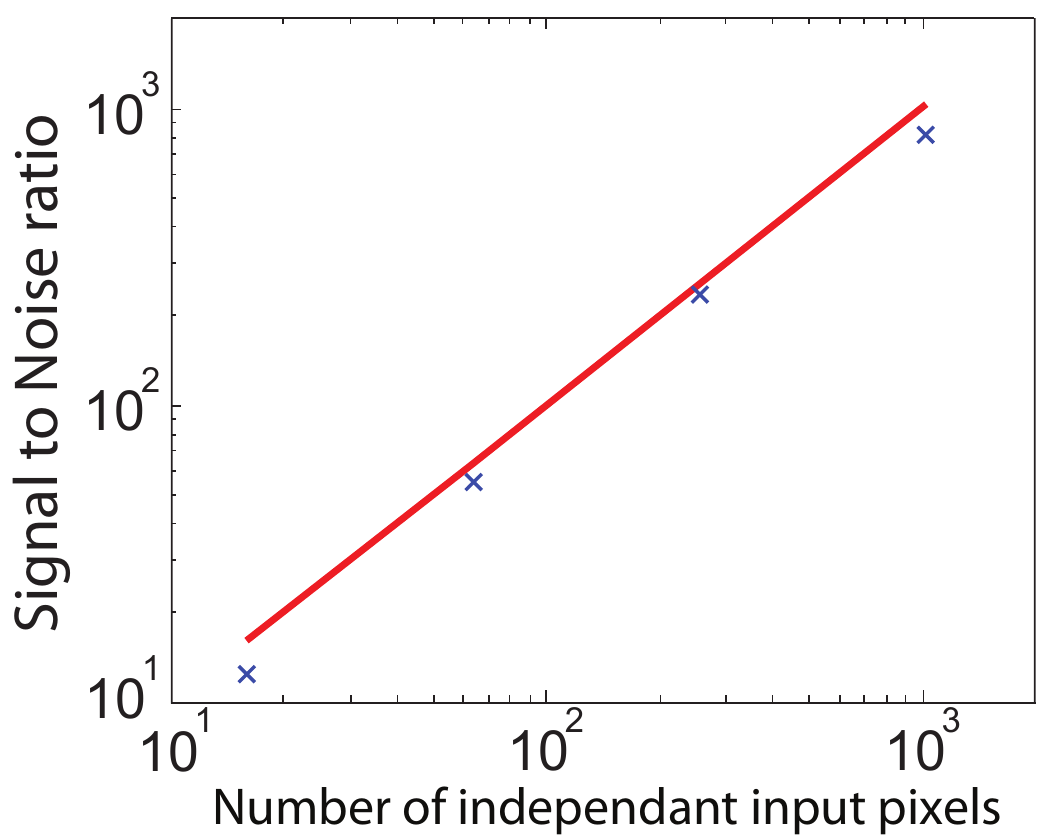}
\caption{ Focusing Signal to Noise Ratio as function of the number of SLM controlled segments $N$. We show in solid line the theoretical prediction $S\!N\!R \approx 0.5 N_{\mathrm{DOF}}$ and with $\times$ the experimental results.} 
\label{SNRvsN}
\end{figure}

To experimentally study the effect of $N_{\mathrm{DOF}}$ over the quality of focusing, we measure the energy signal to noise ratio for a single spot focusing. We represent in figure~\ref{SNRvsN} this experimental ratio as a function of the number of controlled segments $N$ on the SLM and compare it to the theoretical prediction $S\!N\!R \approx 0.5 N_{\mathrm{DOF}}$. Those experimental results correctly fit the theoretical predictions.

\subsection {Comparison with Sequential Optimization}

The first breakthrough using phase-only Spatial Light Modulator for focusing through a disordered medium were reported in~\cite{vellekoop2007focusing}. The technique was a feedback optimization that leads to substantial enhancement of the output intensity at a desired target. We will present the principle of this approach and compare it to techniques requiring the measurement of the TM.

The experimental setup is roughly the same than the one presented in figure~\ref{Setup}. The wavefront phase is controlled onto the SLM on $N$ groups of pixels, and the full output pattern is recorded by the CCD camera. One point target is chosen and the optical field $E^{\mathrm{out}}_m$ on this pixel is the linear combination of the effects of the $N$ segments on the modulator~:

\begin{equation}
E^{\mathrm{out}}_m = \sum_n^N{k_{mn}E_0 \rme^{\rmi\phi_n}}
\label{eqEMosk}
\end{equation}

where $k_{mn}$ are the elements of the TM as defined in equation~\ref{eqE} and $\phi_n$ the phase imposed on the $n^{th}$ pixel of the SLM.

The input amplitude term $E_\mathrm{0}$ is constant for a phase-only spatial light modulator with an homogeneous illumination. The idea is, for each pixel, to test a set of phase values and keep the one that gives the higher intensity on the target output point. The goal of this sequential algorithm is to converge toward a phase conjugation with an efficiency depending on the finite number of phase tested~\cite{vellekoop2007focusing}. Different phase shifting algorithms were studied and their relative efficiencies in presence of time decorrelation are analyzed in~\cite{vellekoop2008phase}, we will not discuss this problem in this paper.

In a single spot focusing experiment for a sufficient number of phase shifts tested, the sequential algorithm and the phase conjugation computed by the knowledge of the TM are equivalent. Nevertheless, both techniques have different advantages and drawbacks. To be able to change the position or the shape of focal pattern, sequential optimization needs a new optimization for each new pattern. This forbids real time applications. On the contrary, once the TM recorded, all information is known and permits to calculate the input wavefront for any output desired pattern, allowing for instance to change the focus location at a video rate.  Nevertheless, sequential optimization presents two strong advantages. With a setup comparable to the one shown on figure~\ref{Setup}, sequential algorithm is more robust to calibration errors than the TM measurement since, even if the exact values of the phase shifts tested are unknown, the algorithm automatically selects the best phase mask. Another asset of this technique is that no interferometry measurement is needed. Since only the intensity is recorded, it allows experiments using incoherent phenomenon such as focusing optical intensity on a fluorescence probe inside a disordered material~\cite{vellekoop2007demixing}.

\section{Noise and Reconstruction Operator}
\label{SecNoise}
Experimentally, one cannot perform a measurement without noise. In our setup, the main noise sources are the laser fluctuation, the CCD readout noise and the residual amplitude modulation. Because of the error made on the TM measurement, one has to find, for focusing or image detection, reconstructions operators very resilient to experimental noise. These operators only give an estimation of the solution. From now on, we will make the distinction between the two sources of perturbation on the reconstructed signal : the experimental noise inherent to the measurement errors, and the reconstruction noise, inherent to the reconstruction operator. We will discuss the influences of those two noises on the performance of the technique used. Due to spatial reciprocity, focusing and image detection are equivalent. We will thus concentrate in this section on the case of image detection.

Inverse problems are widely commons in various fields of physics. Optical image transmission through a random system is the exact analog of Multiple-Input Multiple-Output (MIMO) information transmission through an unknown complex environment. This has been broadly studied in the last decades since the emergence of wireless communications~\cite{introPaulraj}. Similar problems have also been studied in optical tomography in incoherent~\cite{arridge1999optical} and coherent regime~\cite{maire2009experimental}. In a more general way, inverse problems has been widely study theoretically for more than fifty years. A. Tikhonov~\cite{tikhonov1963solution} proposed a solution for ill-posed system. Tikhonov regularization method was then broadly used in various domains to approximate the solutions of inverse problems in the presence of noise.

\subsection{Experimental Noise}
\label{expNoise}

Prospecting an optimal operator for image reconstruction, the first straightforward option is the use of inverse matrix $K_{\mathrm{obs}}^{-1}$ for a symmetric problem ($N = M$) since $K_{\mathrm{obs}}^{-1}K_{\mathrm{obs}} = I$ where I is the identity matrix. Inversion can be extended when $N \neq M$ with the pseudo-inverse matrix $\left[K_{\mathrm{obs}}^{\dag}.K_{\mathrm{obs}}\right]^{-1}K_{\mathrm{obs}}^{\dag}$, noted $K_{\mathrm{obs}}^{-1}$. This operator provides a perfect focusing but is unfortunately very unstable in presence of noise. Indeed, the singular values of $K_{\mathrm{obs}}^{-1}$ are the inverse ones of $K_{\mathrm{obs}}$. In other words, inverse operators tend to increase the energy sent through otherwise low throughput channels. This way, it normalizes the energy coming out from all channels. Singular values of $K_{\mathrm{obs}}$ below noise level have the strongest contributions in $K_{\mathrm{obs}}^{-1}$ but are aberrant. The reconstructed image is hence dominated by noise and bears no similarity with the input one. Another standard operator for image reconstruction or communication is Time Reversal. This technique has been studied in acoustics~\cite{FinkPhysTo97} and electromagnetic waves~\cite{lerosey2007focusing}. Being a matched filter~\cite{tanter2000time}, Time Reversal is known to be stable regarding noise level. Unlike inversion, it takes advantage of the strong singular values by forcing energy in the channels of maximum transmission. This way, singular values below noise level carry few energy and do not degrade reconstruction. Its monochromatic counterpart the Phase Conjugation uses the operator $K_{\mathrm{obs}}^{\dag}$ (the operator used in~\ref{expPC} for focusing and target detection). $K_{\mathrm{obs}}^{\dag}K_{\mathrm{obs}}$ has a strong diagonal but the rest of it is not null. We will see later that this implies that the fidelity of the reconstruction is not perfect and depends on the complexity of the image to transmit.

An intermediate approach is to use a Mean Square Optimized operator. This operator, that we will call MSO an denotes $W$ is the Tikhonov regularization~\cite{tikhonov1963solution} for linear perturbated systems mentioned earlier. It takes the experimental noise into account minimizing transmission errors, estimated by the expected value $E\left\{\left[W.E^{\mathrm{out}}-E^{\mathrm{in}}\right]\left[W.E^{\mathrm{out}}-E^{\mathrm{in}}\right]^{\dag}\right\}$.  For an experimental noise of variance $No_{\sigma}$ on the output pixels, $W$ reads~:

\begin{equation}
W = \left[K_{\mathrm{obs}}^{\dag}.K_{\mathrm{obs}} + No_{\sigma}.I \right]^{-1}K_{\mathrm{obs}}^{\dag}.
\label{MSO}
\end{equation}

This operator stabilizes inversion through the addition of a constraint depending on the noise level. This operator is intermediate between the two previous one. If $No_{\sigma}=0$, $W$ reduces to the inverse matrix $K_{\mathrm{obs}}^{-1}$, which is optimal in this configuration, while for a very high noise level it becomes proportional to the transpose conjugate matrix $K_{\mathrm{obs}}^{\dag}$, the phase conjugation operator. For an intermediate noise level, the channels of transmission  much greater than the noise level are used 'like in inversion' operator and channels much below noise level are used 'like in phase conjugation'.

\subsection{Reconstruction Noise}

The reconstruction noise is the perturbation brought by the reconstruction operator, even without measurement noise. This noise is intrinsic to the techniques used. The only operator to provide a perfect reconstruction, \textit{i.e.} without any reconstruction noise, is pseudo-inversion (or inversion for $M = N$). Nevertheless, this operator is not usable in a general case since it is very sensitive to experimental noise. Phase conjugation is stable in presence of experimental noise but creates perturbations of the reconstructed signal. As seen in section~\ref{NDOF}, the signal to noise ratio for a single focal is proportional $N_{\mathrm{DOF}}$. The same goes for imaging a single spot, the target to detect generates a noise proportional to $1/N_{\mathrm{DOF}}$ on the rest of the image. For multiple targets, each target detection adds a perturbation for the reconstruction of the others, thus the intensity signal to noise ratio is approximately given by $N_\mathrm{f}/N_{\mathrm{DOF}}$ where $N_\mathrm{f}$ is the number of targets to detect. In other words, the fidelity of the reconstruction with Phase Conjugation is not perfect and depends on the complexity of the image to transmit. This limitation forbids the reconstruction of a complex image (with $N_\mathrm{f} \approx N$) by Phase Conjugation through a symmetric system (with $N = M$) since the reconstruction noise is of the same order of magnitude as the useful signal.

For any reconstruction operator $Op$, if $K$ is known exactly (not affected by the experimental noise), the mean normalized reconstruction noise $N_{\mathrm{rec}}$ for a single target detection can be estimated by :

\begin{equation}
N_{\mathrm{rec}} = \frac{\left\langle |Op \times K|_{mn}^2\right\rangle_{m\neq n}}{\left\langle |Op \times K|_{mn}^2\right\rangle_{m = n}}
\end{equation}

\begin{figure}[ht]
\center
\includegraphics[width=0.6\textwidth]{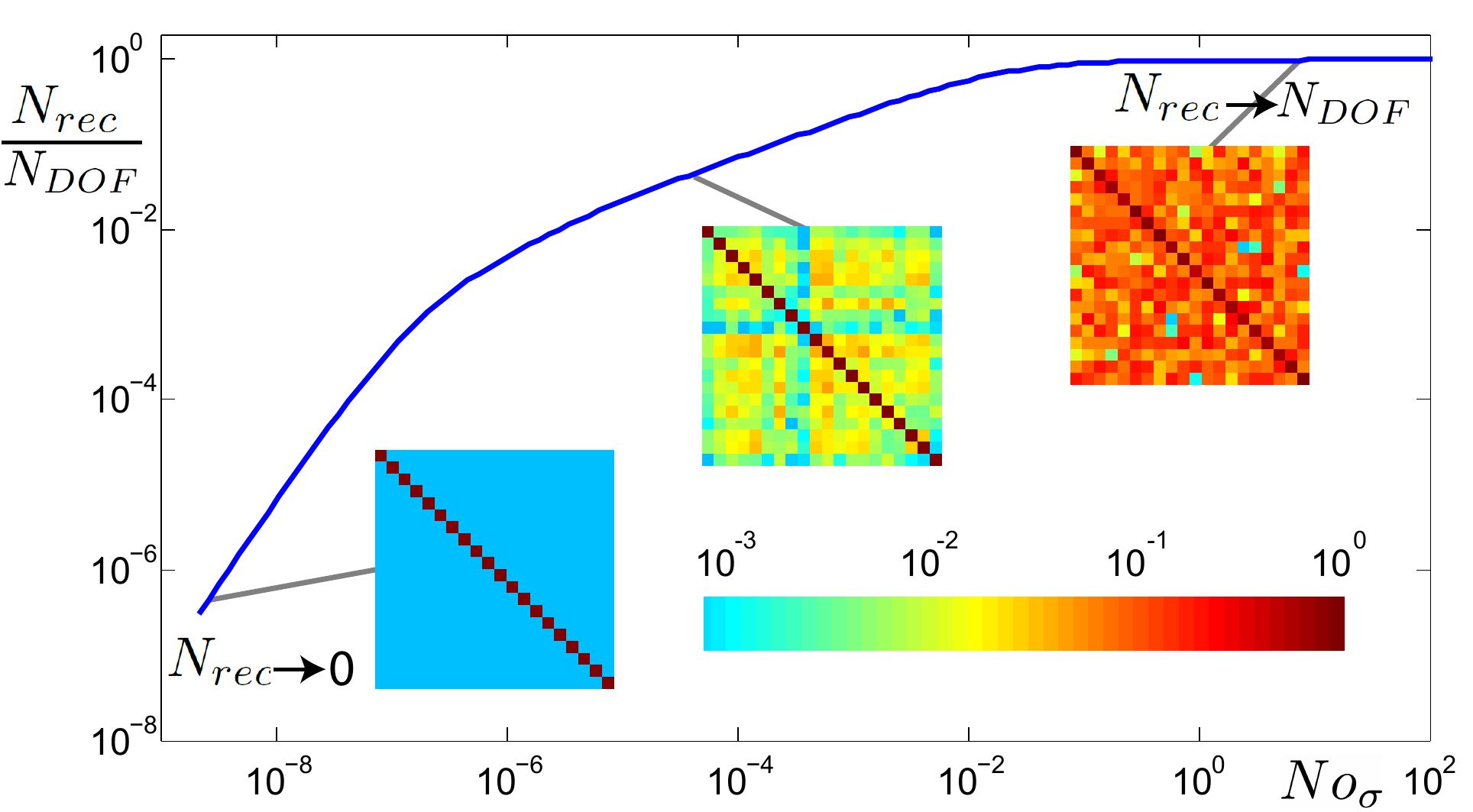}
\caption{Reconstruction noise as a function of $No_{\sigma}$. Results are obtained in simulation for a TM modelized by a 20 by 20 random matrix of Gaussian statistics. $No_{\sigma}$ is normalized by the total transmission energy $\sum_k{\lambda^2_{k}}$. In inset we show the values of $|W\times K|^2$ with a logarithmic scale.} 
\label{NRvsNoise}
\end{figure}

$N_{\mathrm{rec}}$ is define here by the ratio between the mean intensity $\left\langle |Op \times K|_{mn}^2\right\rangle_{m\neq n}$ brought by the reconstruction of a target outside its position and the mean intensity $\left\langle |Op \times K|_{mn}^2\right\rangle_{m = n}$ at the target position. The average is made over all input target positions. Graphically, $N_{\mathrm{rec}}$ is the inverse of the signal to noise ratio defined by the mean of the intensity of the diagonal of $Op \times K$ over the mean intensity elsewhere.\\
To study the reconstruction noise brought by MSO operator for different values of $No_{\sigma}$ we show in figure~\ref{NRvsNoise} the simulation results for the reconstruction noise as a function of the $No_{\sigma}$. It is important to understand that, for a given $No_{\sigma}$, $W$ is the ideal operator for an experimental energy noise level of $No_{\sigma}$. In that study, the simulated TM $K$ is noise free. We see that for $No_{\sigma} = 0$ (\textit{i.e.} for inversion operator), the reconstruction noise $No_{\mathrm{rec}}$ is null and for $No_{\sigma}\rightarrow \infty$ (\textit{i.e.} for Phase Conjugation operator) $No_{\mathrm{rec}}$ tends to $N_{\mathrm{DOF}}$ (here $N_{\mathrm{DOF}} = M =N$). This behavior is obvious when looking at the diagonal of $|W\times K|^2$ compared to the others values. 

\subsection{Toward Optimal Reconstruction}
\label{optimalRec}

In order to detect an input image with the best accuracy, one has to find the good balance between the sensitivity to experimental noise and the influence of the reconstruction noise on the recovered image. The mathematical optimum operator for those constrains is the MSO operator introduced in equation~\ref{MSO}. To lower the reconstruction noise, one can only increase the parameter $N_{\mathrm{DOF}}$. We investigated two possibilities: averaging, and increasing the number of output modes $M$.

A possible way to virtually increase $N_{\mathrm{DOF}}$ is to perform the transmission of the same image through different realizations of disorder. With several realizations, the reconstruction noise, depending on the TM, will be different and averaging will lower perturbations on the final image. It is the monochromatic equivalent of using broadband signals, which takes advantage of temporal degree of freedom~\cite{lemoult2009} where uncorrelated frequencies are used to increase $N_{\mathrm{DOF}}$. Iterating with different physical realizations of disorder is restrictive since it requires to actively move or change the scattering sample. If we are only interesting in detecting an amplitude object, an other way to change the effective TM seen by the object is to illuminate it with different phase wavefronts. It is formally equivalent to transmitting the same image through different channels as if the image propagated through different realizations of disorder. For any amplitude object $E^{\mathrm{obj}} \quad \text{with} \quad e^{\mathrm{obj}}_m \in \left[0,1\right]$, the effective incident field on the sample is $S'_{\phi}.E^{\mathrm{obj}}$ where $S'_{\phi}$ is a diagonal matrix containing only phase terms. The output complex field pattern for a given TM $K$ reads~:

\begin{equation}
E_{\mathrm{out}} = K.S'_{\phi}.E_{\mathrm{obj}} = K'.E_{\mathrm{obj}}
\label{Sphi}
\end{equation}

With $K'= K.S'_{\phi}$. Physically, it is as if the same object $E_{\mathrm{obj}}$ was transmitted through a different part of the medium. Another simple way to see it is that by changing $S'_{\phi}$ (\textit{i.e.} changing the 'virtual realization') we change the projection of $E_{\mathrm{obj}}$ on the input singular modes. Thus the same image propagates through different channels. For a given phase illumination, the amplitude object is estimate by $E_{\mathrm{img}}$ :

\begin{equation}
E_{\mathrm{img}} = |W.E_{\mathrm{out}}| = |W.K.S'_{\phi}.E_{\mathrm{obj}}| = |W.K'.E_{\mathrm{obj}}|
\label{EimgSphi}
\end{equation}

\begin{figure}[ht]
\center
\includegraphics[width=0.7\textwidth]{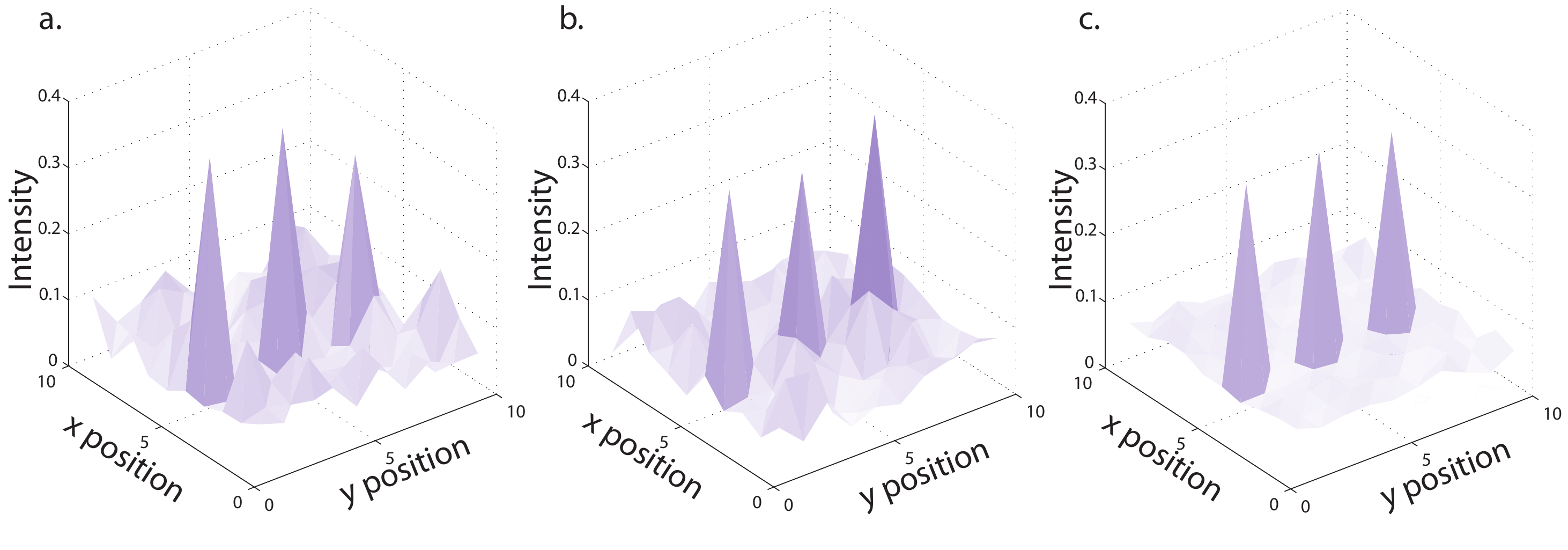}
\caption{Comparisons of reconstruction simulations of a simple object for two different 'virtual realizations' (a. and b.) and the reconstruction averaged over over 10 'virtual realizations' (c.). Simulations are made for a 100 by 100 random matrix of independent elements of Gaussian statistics. The amplitude object consists of three points of amplitude one and the rest of the image to zero.}
\label{VRfoc}
\end{figure}

With $W$ the optimal MSO operator corresponding to the TM $K$ for a given noise level. One now has to change the illumination wavefront $S'_{\phi}$ and average the different $E_{\mathrm{img}}$ obtained to lower the reconstruction noise. To illustrate this effect we show in figure~\ref{VRfoc} simulation results for the reconstruction of a simple amplitude object using two different phase illuminations and averaging over 10 different illuminations. The matrix is know exactly in the simulation so all perturbations come from reconstruction noise. We see that the two first realizations reproduce the amplitude object with different noises with amplitude of the same order of magnitude. Averaging over multiple illuminations significantly reduces the noise background.

Since our SLM is a phase only modulator, in order to simulate amplitude objects, we generate what we call 'virtual objects'. We use different combinations of random phase masks to generate the same virtual amplitude object $E^{\mathrm{obj}}$ by subtracting two phase objects. From any phase mask $E^{(1)}_{\mathrm{phase}}$ one could generate a second mask $E^{(2)}_{\mathrm{phase}}$ where the phase of the $m^{th}$ pixel is shifted by $s^{\mathrm{obj}}_m.\pi$. The optical complex amplitude of elements of the second mask are linked to those of the first mask by :

\begin{equation}
e^{(2)}_m = e^{(1)}_m.\rme^{\rmi s^{\mathrm{obj}}_m\pi}
\end{equation}

with $e^{(j)}_m$ the $m^{th}$ element of $E^{(j)}_{phase}$. We have the relation between the two masks and the amplitude of the 'virtual object' :

\begin{equation}
|E^{(2)}_{\mathrm{phase}} - E^{(1)}_{\mathrm{phase}}| = 2.\sin{\left(E_{\mathrm{obj}}\pi/2\right)}
\end{equation}

it is important to note that the norm of $E^{(2)}_{\mathrm{phase}} - E^{(1)}_{\mathrm{phase}}$ is fixed and controlled whereas its phase is random, uncontrolled and differs from a 'virtual realization' to another. The term $E_{\mathrm{amp}} = |E^{(2)}_{\mathrm{phase}} - E^{(1)}_{\mathrm{phase}}|$ can be estimated by : 

\begin{equation}
E_{\mathrm{img}} = |W.(E^{(2)}_{\mathrm{out}} - E^{(1)}_{\mathrm{out}})| 
\end{equation}

where $E^{(1)}_{\mathrm{out}}$ (resp. $E^{(2)}_{\mathrm{out}}$) is the complex amplitude of the output speckle resulting from the display of the mask $E^{(1)}_{\mathrm{phase}}$ (resp.  $E^{(2)}_{\mathrm{phase}}$). From a given set of random phase masks, we can define in the same way as in equation~\ref{Sphi} a diagonal matrix $S'_{\phi}$ containing only phase terms. This term differs from a 'virtual realization' to another. This way, the reconstruct image reads~:

\begin{equation}
E_{\mathrm{img}} = |W.S'_{\phi}.E_{\mathrm{amp}}| 
\end{equation}

This expression is formally equivalent to equation~\ref{EimgSphi}, thus generating 'virtual objects' using random phase masks is mathematically equivalent to illuminating a real object by random phase-only wavefront.

\subsection{Optimal Reconstruction in Presence of Ballistic Light}
\label{BallSec}

We experimentally used this technique to virtually increase $N_{\mathrm{DOF}}$, and we average the results to lower the reconstruction noise. The complex amplitude image to detect is a grayscale 32 by 32 pixels flower image. The system is symmetric with $N = M = 1024$. The first step to reconstruct the input image with the best accuracy is to find the variance of the experimental noise $No_{\sigma}$ in order to obtain the optimal MSO operator. To that end, we numerically iterate the reconstruction operation for different values of the constrain in the MSO operator (equation~\ref{MSO}) and keep the one maximizing correlation between the reconstructed image and the image sent, hence obtaining an estimation of the experimental noise level. This step is made \textit{a priori} knowing the image sent but could be performed by any technique leading to an accurate estimation of the noise level.  We detailed in~\cite{popoff2010image} the results in a purely diffuse system where ballistic contributions have no measurable effect. We study here the results in presence of quantitative ballistic contributions.

In previous experiments, the focal planes of the two objectives are not the same. If the two focal planes coincide, if we remove the sample, an incident \textbf{k}-vector matches an output \textbf{k}-vector. Thus a pixel of the SLM corresponds to a pixel of the CCD. In the same configuration with a scattering medium, if the part of the energy ballistically transmitted is important enough, this ballistic part can be isolated in the TM. 

\begin{figure}[ht]
\center
\includegraphics[width=0.55\textwidth]{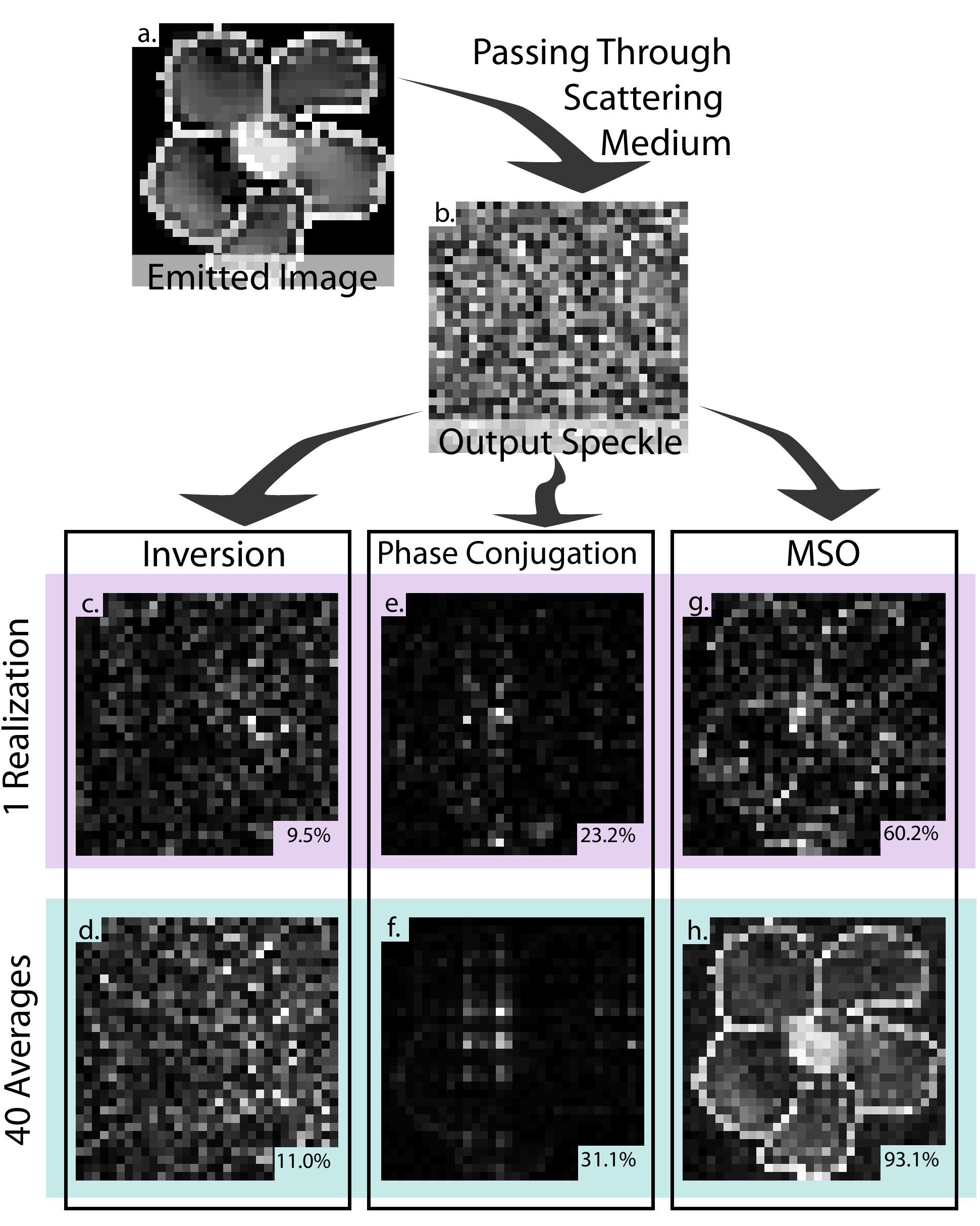}
\caption{Comparisons of the reconstruction methods in presence of ballistic contributions. \textbf{a.} initial gray scale object and \textbf{b.} a typical output speckle figure after the opaque medium. \textbf{c.}, \textbf{e.} and \textbf{g.} are experimental images obtained with one realization using respectively inversion, phase conjugation and MSO operator, \textbf{d.}, \textbf{f.} and \textbf{h.} are experimental images averaging over 40 'virtual realizations' using respectively inversion, phase conjugation and MSO operator. Values in insets are the correlation with the object \textbf{a.}.}
\label{MSO_ball}
\end{figure}

To observe those contributions and study the accuracy of the reconstruction methods in that particular case, we experimentally use a thinner and less homogeneous area of the medium with a symmetric setup ($M = N$) where input and output segments are approximately the same size. We test the different reconstruction methods presented in~\ref{optimalRec} with one realization and averaging over 40 'virtual realizations' with random phase masks. Results are shown in figure~\ref{MSO_ball}. As expected, we see that the inversion operation does not allow image reconstruction, even with averaging, whereas optimal MSO allows a 93\% correlation for 40 averaging (and a 60\% correlation for one realization). Phase conjugation which is supposed to be stable in presence of noise gives a weak correlation of 31\% with averaging.

\begin{figure}[ht]
\center
\includegraphics[width=0.55\textwidth]{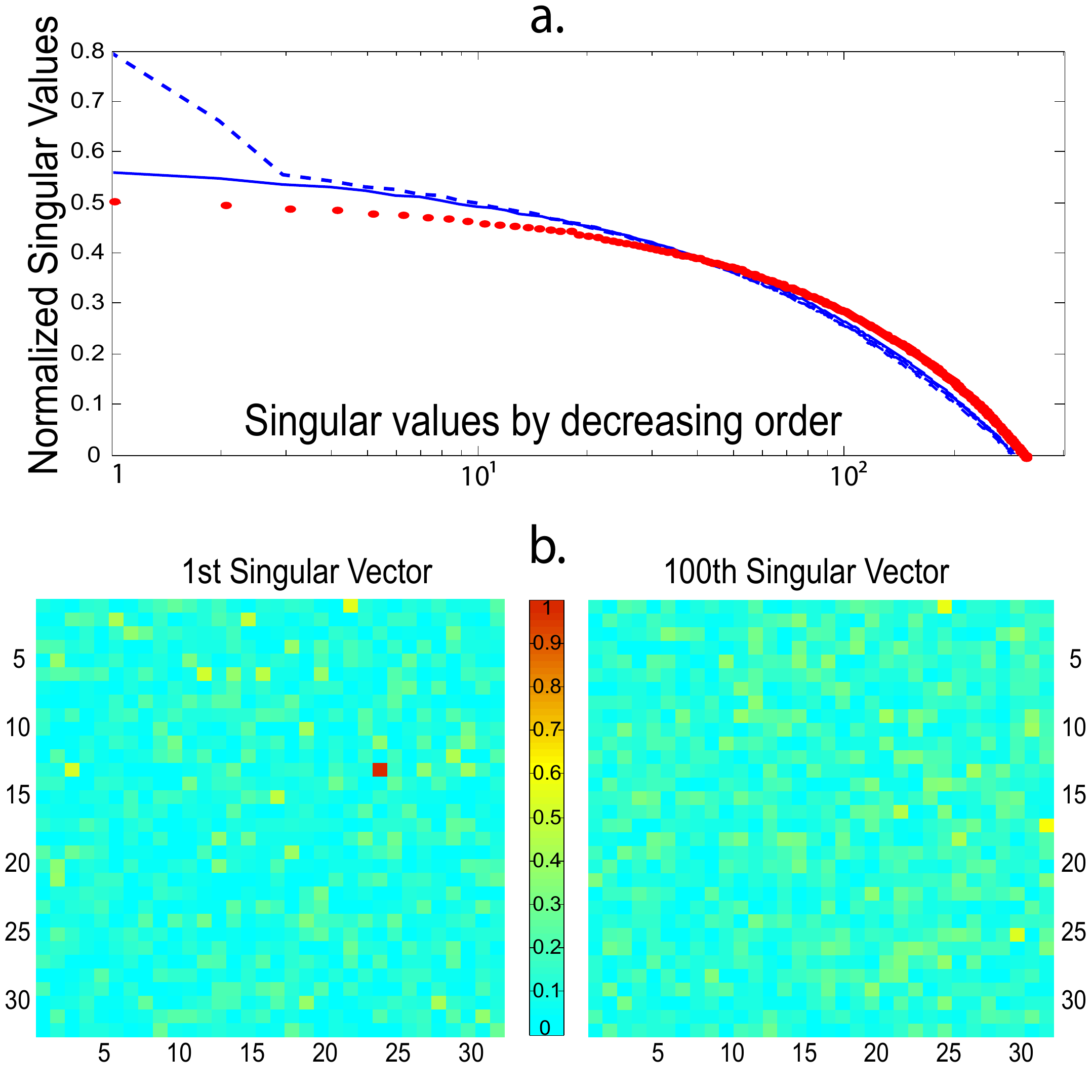}
\caption{ (a.) normalized singular values $\widetilde{\lambda}$ sorted by decreasing order for a thin non homogeneous region of the sample with experimental conditions sensitive to ballistic waves (dash line) and for a thicker region of the medium with the two focal planes objectives separated by 0.3 mm (solid line). With $\bullet$ we show the normalized singular values of a simulated matrix of independent Gaussian elements following the 'quarter circle law'.(b.) amplitude of the singular vector corresponding to the maximum singular value (left) and to the 100$^{th}$ singular value (right). Both share the same amplitude scale.}
\label{BallVec}
\end{figure}

Those results are similar to the ones obtained when we measure the TM of a system governed by multiple scattering~\cite{popoff2010image} for inversion (11\% correlation with averaging) and for MSO operator (94\% correlation with averaging) but are very different for phase conjugation (76\% correlation with averaging). We see in figure~\ref{MSO_ball}(e.) and (f.) that most of the energy of the reconstructed field is localized on very few points. In presence of strong enough ballistic components, the direct paths associated give rise to channels of high transmission linking one input \textbf{k}-vector to the same output \textbf{k}-vector. Figure \ref{BallVec}(a.) shows normalized singular values for both experiments and for a simulated random matrix following the quarter circle law. For the 'ballistic' experiment, two high singular values can not be explained with a TM governed by multiple scattering only. We show for comparison in figure \ref{BallVec}(b.) the spatial aspect of the input singular vectors corresponding to the first and the 100$^{th}$ highest singular values, which should not be affected by balistic light. The 100$^{th}$ singular vector is spatially randomly distributed in energy (as expected for random multiple scattering) whereas the first singular vector mostly contains energy in one \textbf{k}-vector. This confirms that, in this particular case, the channel of maximum transmission is mostly dominated by the ballistic contribution.

Phase conjugation maximizes the energy transmission in channels of maximum transmission~\cite{tanter2000time}. Therefore, ballistic high singular values contributions will be predominant in phase conjugation, regardless of the image $E_{\mathrm{obj}}$. Since the ballistic singular vectors are not homogeneously spatially distributed in energy, they will not efficiently contribute to an arbitrary image reconstruction. MSO is not affected since lower the weight of channels which do not efficiently contribute to the image reconstruction, including the ballistic ones.

\subsection{Pseudo-Inversion}

We showed in section~\ref{model} that when we set the output pixel size to fit the output coherence length, the number of degrees of freedom for detection is equal to the number $M$ of segments read on the CCD. This way a second approach to increase $N_{\mathrm{DOF}}$ and thus the quality of the reconstructed image is to enlarge the output observation window. An important advantage is that the limiting time in our experiment is proportional to the number of input basis vectors $N$, thus, in contrast with a focusing experiment where $N_{\mathrm{DOF}}$ is equal~\cite{vellekoop2007focusing} to $N$, increasing the degrees of freedom for image reconstruction does not increase the measurement time. Another consequence of increasing the size of the image recorded, that is, of increasing the asymmetric ratio $\gamma = M/N \geq 1$, is that the statistical properties of the matrix are deeply modified. We saw in section~\ref{MSTM} that increasing $\gamma$ the range of the normalized singular values decreases~\cite{marcenko1967distribution, sprik2008eigenvalue}. A direct consequence is that the smallest non-zero singular value increases and reads $\lambda^0_{\gamma} = (1-\sqrt{1/\gamma})$. Controlling less information on the input than measured on the output results in an averaging effect. An interesting consequence is that for a reasonable experimental noise level, one can find a value of $\gamma$ for which all singular values (and thus all channels energy transmission) are above this noise level. In such a situation, the TM recorded is barely sensitive to the experimental noise and since no singular values are drowned in the noise, pseudo-inverse operator can be efficiently used.

\begin{figure}[ht]
\center
\includegraphics[width=0.55\textwidth]{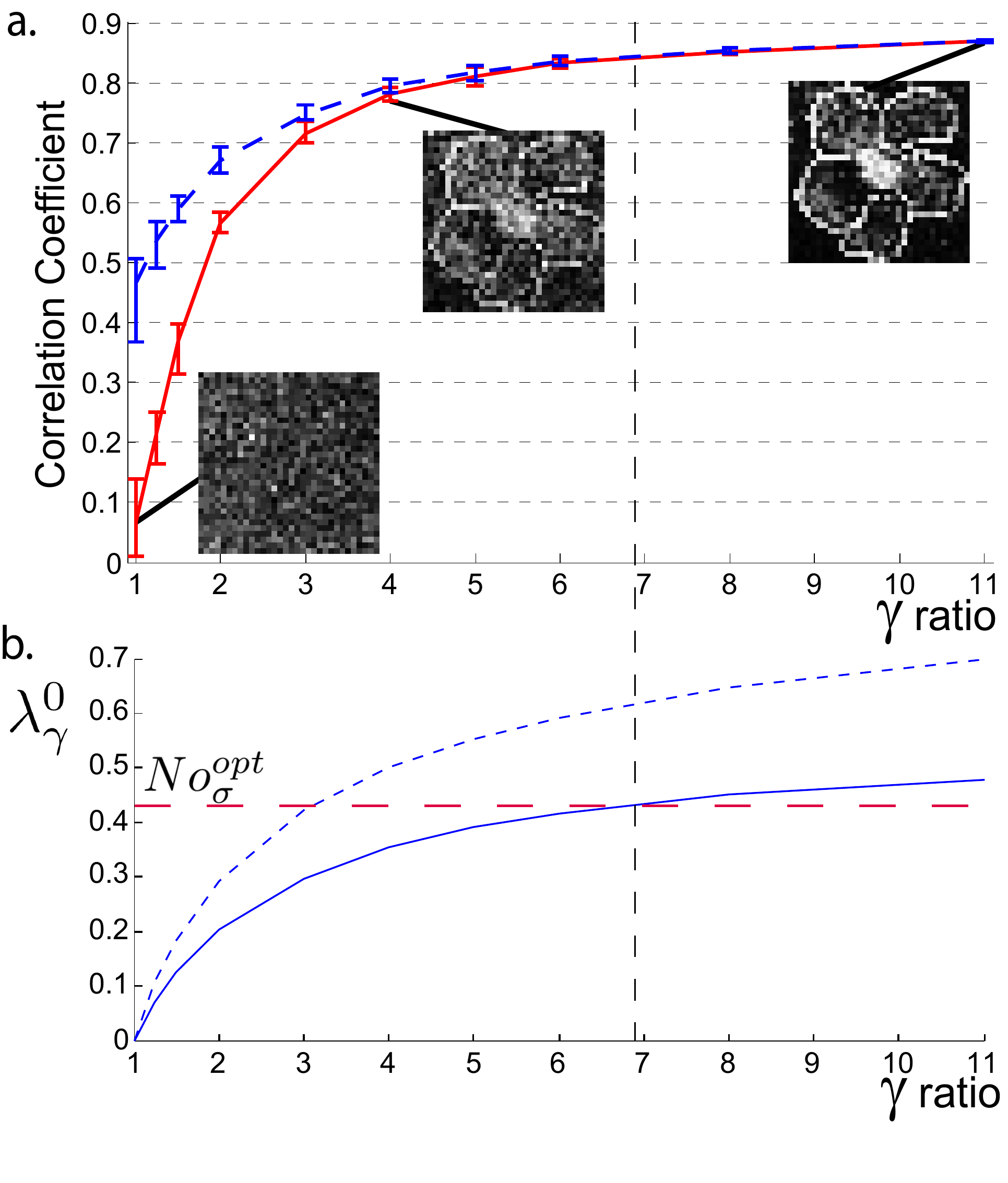}
\caption{ Influence of the number of output detection modes. \textbf{a.} correlation coefficient between $E_{\mathrm{img}}$ and $E_{\mathrm{obj}}$ as a function of the asymmetric ratio $\gamma = M/N$ for MSO (dashed line) and for pseudo-inversion (solid line), without any averaging. Error bars correspond to the dispersion of the results over 10 realizations. \textbf{b.} Experimental (solid line) and Marcenko-Pastur~\cite{marcenko1967distribution} predictions (dashed line) for the minimum normalized singular value as a function of $\gamma$. The horizontal line shows the experimental noise level $No_{\sigma}^{\mathrm{opt}}$. } 
\label{Gamma}
\end{figure}

To highlight this effect, we experimentally measure the TM of our sample for different values of $\gamma \geq 1$. For each TM we tested a single realization of the reconstruction of a complex image using optimal MSO operator and pseudo-inverse operator. The results are shown in figure~\ref{Gamma}(a.). For $\gamma = 1$ we are in agreement with previous results since optimal MSO gives a partially accurate image when results of inversion are totally drown in noise. Increasing $\gamma$ and thus $N_{\mathrm{DOF}}$ strongly improves the quality of the reconstructed image for both techniques and reaches a $>85\%$ fidelity for the largest value of $\gamma=11$, without any averaging. As expected, the curves of fidelity reconstruction as function of $\gamma$ for both techniques become identical when the value of $\lambda^0_{\gamma}$ reaches noise level (figure~\ref{Gamma}(b.)). Above this value, pseudo-inversion and optimal  MSO are totally equivalent.
One notice that experimental $\lambda^0_{\gamma}$ are always smaller than their theoretical predictions. We explain this deviation by the correlation induced in the matrix by the reference pattern $|S_{\mathrm{ref}}|$ (see section~\ref{MSTM}).

We saw in this part that increasing the degrees of freedom and using an appropriate operator that takes noise into account permits to drastically enhance the reconstruction accuracy of an image detection. Similar results can be obtain for focusing or for output wavefront shaping with the same techniques. Such experiments are more difficult to carry out since they require amplitude and phase modulation. Increasing $\gamma=N/M$ and thus $N$ in that case also presents the drawback to significantly increase the measurement time.

\section{Memory Effect}
\label{BallMem}
\label{MemoryEffect}

To define the statistical properties of the TM of multiple scattering media in part~\ref{MSTM} we did not take into account weak correlation effects that could break the hypothesis of independent elements of the TM. The strongest (first order) of those effects that can be recorded for system dominated by multiple-scattering (no ballistic contribution) in the diffusive regime is the so-called Memory Effect (ME). It brings correlation with characteristic length $L_{\mathrm{me}}$ which depends, in transmission, on the geometry of the observation system~\cite{PhysRevLett.61.2328}. If the speckle grain size $L_{\mathrm{speckle}}$ is of the same order of magnitude or smaller than the ME range $L_{\mathrm{me}}$, ME brings additional correlations in the TM.

We define $L_{\mathrm{speckle}}$ as the full width at half maximum of the autocorrelation function of a speckle intensity pattern and $L_{\mathrm{me}}$ as the attenuation length of the ME. Both length are defined for a given observation plane.

The speckle size $L_{\mathrm{speckle}}$ depends on the distance between the back of the sample and the observation plane $D$, the width of the illumination area $W$ and the wavenumber $k$. It reads~:

\begin{equation}
L_{\mathrm{speckle}} = \frac{2\pi D}{kW}
\end{equation}

In contrast, $L_{\mathrm{me}}$ depends on $k$, $D$ and the thickness $\mathbf{e}$ of the sample and reads~:

\begin{equation}
L_{\mathrm{me}} = \frac{D}{k\mathbf{e}}
\end{equation}

Therefore, for $W \gg \mathbf{e}$ we have $L_{\mathrm{me}} \gg L_{\mathrm{speckle}}$. This kind of geometry can be convenient since it allows to move a focal spot along a distance greater that $L_{\mathrm{speckle}}$ by just tilting the SLM~\cite{PhysRevLett.106.193905}.\\
We experimentally fixed $D = 1.5 \pm 0.2$ $mm$, $W = 0.5 \pm 0.1$ $mm$ and the sample has a thickness $\mathbf{e} = 80 \pm 25$ $\mu m$. For those conditions, we expect $L_{\mathrm{speckle}} \approx L_{\mathrm{me}} \approx 1.5$ $\mu m$. The TM is measured. Each column of the TM is the complex amplitude response of a plane wave with a given incident angle. One can easily reconstruct the intensity patterns for different incident angles and calculate the correlation function between two patterns~:

\begin{equation}
f_{(n-n'),j} = \left\langle \frac{\left(\left|k_{mn}\right|^2\otimes_m \left|k_{(M-m)n}\right|^2\right)_j}{\sum_m{\left|k_{mn}\right|^2} \times \sum_m{\left|k_{mn'}\right|^2}}\right\rangle_{(n-n')}
\end{equation}

\begin{figure}[ht]
\center
\includegraphics[width=0.55\textwidth]{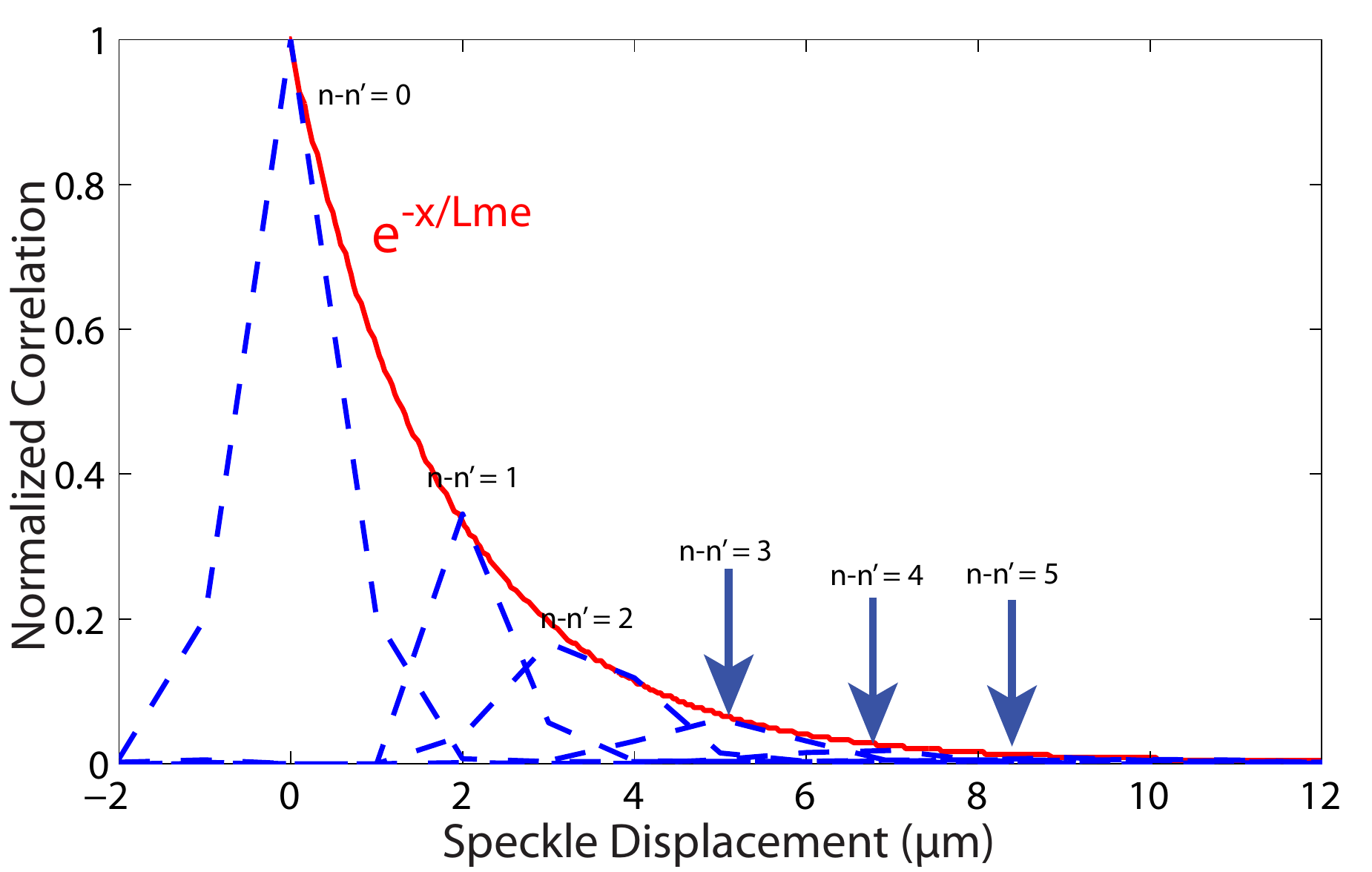}
\caption{Spatial correlation functions of output speckle for a variation of the input incident angle. The correlations functions are drawn as function of the spatial coordinate corresponding to the displacement of the speckle pattern in the observation plane. Arrows show the maximum of correlation functions for the highest displacements. Those peaks are barely observable due to an exponential decay of the maximum correlation value.} 
\label{MECorr}
\end{figure}

Where $\otimes_m$ denotes the spatial convolution over the variable $m$ and $j$ is the variable of the correlation function corresponding to a displacement in pixel in the focal plane. We average results over all combinations of $n,n'$ with $n-n'=$ constant. Results are shown in figure~\ref{MECorr}. It is clear that the output speckle is coherently translated for distances greater than $L_{\mathrm{speckle}}$.

\begin{figure}[ht]
\center
\includegraphics[width=0.55\textwidth]{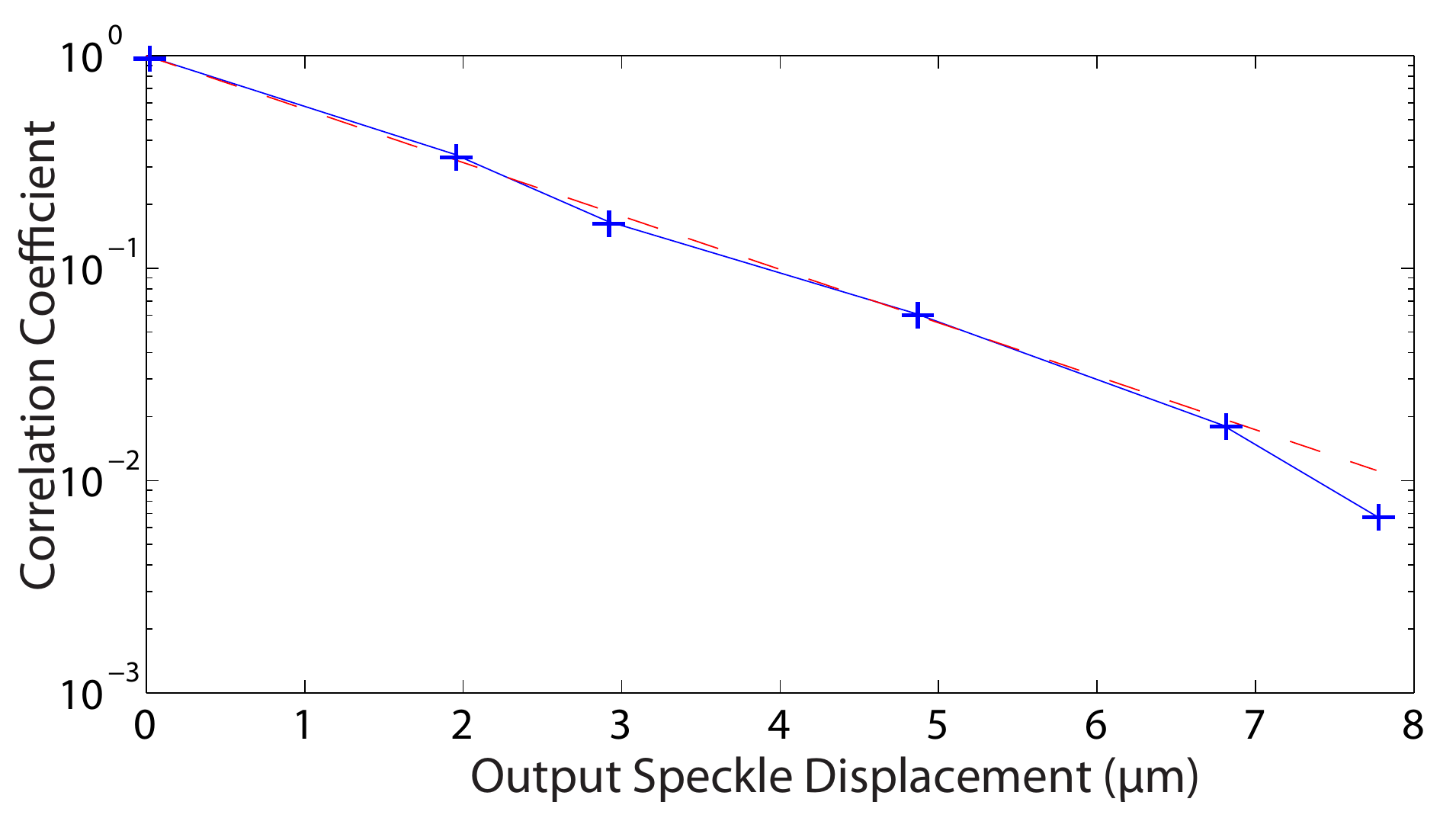}
\caption{Spatial correlation decay of the memory effect. (solid line) experimental curve of correlation coefficient as function of the spatial displacement. (dash line) exponential fit $C(x) = e^{-x/L_{\mathrm{me}}}$ with $L_{\mathrm{me}} = 1.72$ $\mu m$.} 
\label{MECorrMax}
\end{figure}

$L_{\mathrm{speckle}}$ is deduced from the mean autocorrelation function of the speckle pattern $f_{(n-n')=0,j}$ as response to an incident wave plane. We found $L_{\mathrm{speckle}} = 1.3 \pm 0.48$ $\mu m$. We define $C(x)$ the correlation coefficient (maximum of the correlation function) between two output speckles shifted by $x$ in the observation plane.
Exponential decay of the correlation $C(x) \propto e^{-x/L_{me}}$ is generally observed for an open geometry~\cite{PhysRevLett.61.2328}. The measured $C(x)$ (figure~\ref{MECorr} and~\ref{MECorrMax}) fit an exponential decay with $L_{\mathrm{me}} = 1.72 \pm 0.13$ $\mu m$.

We found the expected correlation lengths with a good order of magnitude for autocorrelation (speckle grain size) and memory effect and we were able to qualitatively separate both effects.\\

\section*{Conclusion}
We reinterpreted in this article previous works on optical phase conjugation in term of linear system defined by a monochromatic transmission matrix. Within this formalism, we show how to take advantage of multiple scattering to achieve image transmission and wavefront shaping. We also pointed out how to improve the quality of measurements in the presence of noise, taking benefit of Random Matrix Theory and information theory. We finally showed how to take benefit of the statistical properties of the TM to go back to physical phenomenons.

\ack

This work was made possible by financial support from "Direction Générale de l'Armement" (DGA), Agence Nationale de la Recherche via grant ANR 2010 JCJC ROCOCO, Programme Emergence 2010 from the City of Paris and BQR from Université Pierre et Marie Curie and ESPCI.

\appendix
\section*{Appendix}
\setcounter{section}{1}

\label{appendixA}

To take into account the effect of the reference beam in the effective SNR, we define $\Gamma < 1$ the ratio of the intensity coming from the controlled part over the total intensity. This way, $\Gamma$ is also the ratio of the area of the controlled part of the SLM over the total illuminate area. One way to understand this situation is that $N_{\mathrm{tot}}$ independent segments of the SLM are illuminated and we control only $N = \Gamma.N_{\mathrm{tot}}$ of them. We injected the ratio $\Gamma$ in equation~\ref{incoherentsum} and~\ref{coherentsum}. SNR now reads~:\\

\begin{eqnarray}
&SNR \approx &\frac{\pi/4 N^2 + (N_{\mathrm{tot}}-N)}{N_{\mathrm{tot}}} \\ \nonumber
&=& \frac{\pi/4 N^2 + N(1-\Gamma)}{N/\Gamma} \approx \Gamma \frac{\pi}{4} N^2 \quad \forall N \gg 1 
\end{eqnarray}

For the setup shown in figure~\ref{Setup}, the illuminated area on the SLM represents the circumcircle of the square part that we control. The segments illuminated but not controlled are used to produce the reference pattern. For this geometry, with an illumination disk of radius $R$ we have $\Gamma = \frac{(\sqrt(2).R)^2}{\pi R^2} = \frac{2}{\pi}$. The effective $SNR$ corrected to take into account the reference beam is therefore~:

\begin{equation}
SNR \approx  \left(\frac{\pi}{4}\right)\frac{2}{\pi} N_{\mathrm{DOF}} = 0.5 N_{\mathrm{DOF}}
\end{equation}

\end{document}